\documentclass[aps,prd,a4paper,preprintnumbers,tightenlines,superscriptaddress,
twocolumn,floatfix,showpacs,final]{revtex4-1}
\usepackage{graphicx}
\usepackage{longtable}
\usepackage{amsmath}
\usepackage{amssymb}
\usepackage{subfigure}
\usepackage{hyperref}
\usepackage{dcolumn}
\usepackage{bm}
\usepackage{color}

\newcommand{\be}{\begin{eqnarray}}
\newcommand{\ee}{\end{eqnarray}}
\newcommand{\mpl}{{M_{\rm {pl}}}}

\newcommand{\dd}{\, {\rm d}}

\newcommand{\lsim}{\;\mbox{\raisebox{-0.5ex}{$\stackrel{<}{\scriptstyle{\sim}}$}
}\;}
\newcommand{\rs}{r_{\rm s}}
\newcommand{\ys}{y_{\rm s}}
\newcommand{\pn}{\Phi_{\rm N}}
\newcommand{\pno}{\Phi_{\rm N,0}}
\newcommand{\ver}{\vec{r}}

\newcommand{\gao}{\Gamma_{1,0}}
\newcommand{\thn}{\theta}
\newcommand{\omr}{\omega_R}
\newcommand{\oms}{\omega_{\rm s}}
\allowdisplaybreaks
\begin{document}
\title{Stellar Oscillations in Modified Gravity}
\author{Jeremy Sakstein}
\email{j.a.sakstein@damtp.cam.ac.uk}
\affiliation{Department of Applied Mathematics and Theoretical Physics, Centre
for Mathematical Sciences, Cambridge CB3 0WA,
United Kingdom}

\begin{abstract}
Starting from the equations of modified gravity hydrodynamics, we derive the
equations of motion governing linear, adiabatic, radial perturbations of stars
in scalar-tensor theories. There are two new features: first, the
eigenvalue equation for the period of stellar oscillations is modified such that
the eigenfrequencies are always larger than predicted by General Relativity.
Second, the General Relativity condition for stellar instability is altered so
that the adiabatic index can fall below $4/3$ before unstable modes appear.
Stars are more stable in modified gravity theories. Specialising to the case
of chameleon-like theories, we investigate these effects numerically using both
polytropic Lane-Emden stars and models coming from modified gravity
stellar structure simulations. We find that the change in the oscillation period
of Cepheid star models can be as large as 30\% for order-one matter
couplings and the change in the inferred distance using the period-luminosity
relation can be up to three times larger than if one had only considered the
modified equilibrium structure. We discuss the implications of these results for
recent and up-coming astrophysical tests and estimate that previous methods can
produce new constraints such that the modifications are screened in regions of
Newtonian potential of $\mathcal{O}(10^{-8})$.

\end{abstract}
\maketitle

\section{Introduction}

Modified theories of gravity (MG) have received a lot of attention in the last
decade. Upcoming surveys such as {\scriptsize Euclid}
\cite{Amendola:2012ys} and {\scriptsize LSST} \cite{Ivezic:2008fe} will test
Einstein's theory of General Relativity (GR) on cosmological scales to
incredibly high precision. On the theoretical side, the cosmological constant
problem is driving the study of alternate theories of gravity in an attempt to
explain why the observed vacuum energy density is 120 orders of magnitude
smaller than the quantum field theory prediction (see \cite{Clifton:2011jh} for
a review). Cosmology probes the long-range, infra-red (IR) behaviour of gravity
and so the majority of these theories are IR modifications, often involving new
scalar-degrees of freedom which either drive the acceleration, screen out the
effects of the cosmological constant or include some symmetry such that its
value is technically natural. GR is the unique Lorentz-invariant theory of a
massless spin-2 particle \cite{Weinberg:1965rz} and so any modification of GR 
necessarily introduces at least one new degree of freedom. These could be
decoupled from matter, however, in any generic theory there is no symmetry
forbidding this and so any decoupling is unnatural. If a coupling is present
then one must face the issues of additional or \textit{fifth}-forces that
result whenever a field-gradient is present. GR has been tested to incredibly
high precision over the last century and so one must fine-tune the strength of
these couplings to incredibly small values in order to have a viable theory.
Such small couplings usually render any modifications insignificant in all but
the strongest gravitational fields. For example, any quintessence model of dark
energy requires a mass for the scalar to be of order $H_0$
\cite{Copeland:2006wr} so that the force-range is of the order of the observable
universe. If such a field is then coupled to matter it must pass laboratory
tests, which require the force to be sub-mm \cite{Will:2004nx}. Therefore, the
only way to have a light quintessence-scalar coupled to matter is to tune all of
the couplings to very small values. In general, there is no symmetry protecting
these values and so quantum corrections are expected to induce order-one
corrections. The small value of these couplings is therefore tantamount to
fine-tuning.

There is one caveat to the above argument: all of the tests of GR have been
performed in our local neighbourhood and so there is nothing preventing a theory
which has large fifth-forces active over cosmological scales provided that
these are somehow absent locally. Such theories are said to posses
\textit{screening mechanisms} and there are several examples in the literature
including the chameleon mechanism \cite{Khoury:2003aq,Khoury:2003rn}, the
symmetron mechanism \cite{Hinterbichler:2010es}, the environment-dependent
Damour-Polyakov effect \cite{Brax:2010gi}, the Vainshtein mechanism
\cite{Vainshtein:1972sx} (see also
\cite{Nicolis:2008in,Babichev:2009us,Hinterbichler:2011tt} for applications to
modern theories such as Galileons and massive gravity) and the disformal
mechanism \cite{Koivisto:2012za}. Most of the theories that include a screening
mechanism have the property that the field-profile of the new degrees of freedom
are sourced by the ambient density. They are constructed such that at high
densities either the field's masses are large enough that the force is
short-ranged, the effective coupling to matter becomes negligible or the
field-gradients are suppressed with respect to the gradient of the Newtonian
potential. The exception is the disformal mechanism where field gradients are
sourced by pressure, which is generally negligible for non-relativistic
sources and so no field-gradient is induced.

The problem is then changed from reconciling fifth-forces with local
experiments to distinguishing these theories from GR observationally.
Theories such as Galileons show interesting effects cosmologically through
either the modified background expansion rate, linear perturbation properties or
halo clustering effects
\cite{Barreira:2013jma,Barreira:2013eea,Barreira:2013xea}, however, the same is
not true of chameleons and symmetrons, where the cosmological mass of the field
is always less than $0.1\textrm{Mpc}^{-1}$ \cite{Brax:2012gr,Wang:2012kj},
precluding any effects on linear scales. On non-linear scales, there can
still be some galaxy clustering effects (see, for example
\cite{Hu:2007nk,Li:2011pj,Brax:2012nk,Brax:2013mua} and references therein).
Astrophysical tests of these theories can probe regions of parameter space where
all cosmological signatures vanish and therefore have the potential to constrain
them by several orders of magnitude below what is possible with cosmological
tests. 

Several astrophysical tests of different modified gravity theories have been
investigated, including stellar-structure tests
\cite{Chang:2010xh,Davis:2011qf,Jain:2012tn}, black hole tests
\cite{Hui:2012jb} and galactic dynamics tests
\cite{Jain:2011ji,Lee:2012bm,Vikram:2013uba}. Here, we will be concerned only
with stellar structure. The purpose of this work is to present and investigate
a consistent frame-work for dealing with the equations of modified gravity
scalar fields coupled to hydrodynamics at zeroth- and first-order in
perturbation theory with an aim to looking at both current and new observational
tests. The effects of disformally coupled scalars on the equilibrium structure
of non-relativistic stars are still unknown and a calculation is beyond the
scope of this work. We hence focus on theories where the field is coupled to the
trace of the energy momentum tensor only and the fifth-force is proportional the
the field gradient and the strength of the matter coupling. This includes
chameleon-like and Vainshtein screened theories. Starting from the modified
equations of hydrodynamics, we derive the equations governing the equilibrium
structure of stars (the modified stellar-structure equations) and the evolution
of linearised, radial, adiabatic perturbations of both the field and stellar
structure. We find three new effects in modified gravity:
\begin{enumerate}
 \item When the star is unscreened, the period of oscillation can be reduced by
an $\mathcal{O}(1)$ amount. For Cepheid stars, which are good probes of MG
\cite{Jain:2012tn}, we find that the change in the period can result in
differences from the GR inferred distance measurement of up to three times what
has been previously been calculated using an approximation and therefore that
current bounds can be improved using the same data sets.
\item When the star is unscreened, the star is more stable to these
perturbations. There is a well-known result in stellar astrophysics that when
the adiabatic index falls below $4/3$ the squared-frequency of the
fundamental mode is negative and so there is an unstable mode. We will show
that in MG the index can fall below $4/3$ without any instability. In
MG, there is no universal bound on the critical index for the appearance of the
instability; its precise value depends on the structure and composition of the
star as well as how unscreened it is. 
\item The perturbations of the star can source scalar radiation and vice versa.
\end{enumerate}
The first two effects require the star to be unscreened while the third does
not. Vainshtein-screened theories are too efficient to leave any star
unscreened and so it is only the third effect that may be used to probe this
theory. In this work, we will only investigate the first two effects and so
for this reason, we will only deal explicitly with chameleon-like theories,
although we will present the general derivation and comment briefly on the
application to Galileon-like theories. 

Chameleon-like theories screen according to the local Newtonian potential;
objects where this is lower than a universal model parameter (to be defined
below) are unscreened whereas they are screened when the converse is true. One
is then led to look for MG effects in the most under-dense regions of the
universe. In terms of the Newtonian potential, these are dwarf galaxies. Dwarf
galaxies located in clusters are screened by the Newtonian potential of their
neighbours whilst isolated dwarf galaxies may show MG effects. Recently, MG
N-body codes have been combined with SDSS data to produce a screening map of the
nearby universe \cite{Cabre:2012tq}. This has allowed dwarf galaxy tests of MG
using current data to be performed for the first
time \cite{Jain:2012tn,Vikram:2013uba}.

In unscreened dwarf galaxies located in cosmological voids, main- and
post-main-sequence stars may become partially unscreened leading to strong
fifth-forces in their outer layers. The equilibrium structure of stars in
chameleon-like theories has been well studied \cite{Chang:2010xh,Davis:2011qf}. 
These stars must provide a stronger outward pressure gradient in order to combat
these additional forces than they would had they been GR stars. The extra energy
per unit time required to source this gradient comes from an increased rate of
nuclear burning in the core. A star under the influence of MG is therefore more
luminous, more compact and has a higher effective temperature than its GR
doppelg\"{a}nger. A modified version of the publicly available code {\scriptsize
MESA}, which can simulate the structure and evolution of stars in these theories
was presented in \cite{Davis:2011qf} and has become a powerful tool in the
quantitative prediction of astrophysical effects. At the level of perturbations,
Cepheid-variable stars pulsate at a higher frequency and so obey a different
period-luminosity (P-L) relation to the GR one. Using a simple approximation to
find the new period, \cite{Jain:2012tn} have used {\scriptsize MESA} models to
place the strongest constraints in the literature to date. The MG frequencies
can be numerically computed by solving the new equations coming from modified
gravity hydrodynamics and one aim of this work is to compare the new MG
predictions with this approximation. The others are to investigate the
stability of stars in these theories as well as the implications for current
and possible future astrophysical tests. 

This paper is then organised as follows: in section \ref{sec:mod_grav}, we
present the action for chameleon-like theories and outline some of the essential
features. We then examine the screening mechanism and present the two
model-independent parameters that can be used to fully specify the effects of MG
on the structure and evolution of stars. We then give a brief introduction to
astrophysical tests of modified gravity and outline the current constraints. In
section \ref{sec:MGhydro} we derive the equations of modified gravity
hydrodynamics. At zeroth-order, these give the modified equations of stellar
structure. At first-order, we derive the equations of motion describing the
coupled linear, radial, adiabatic perturbations of both the field and radial
displacements. In section \ref{sec:MLAWE} we specialise to the case where the
scalar-radiation is negligible and elucidate the properties of the new equation
of motion for stellar perturbations. In section \ref{sec:stellarstability} we
show how the criterion for the star to be stable to these perturbations is
altered in and argue that stars are more stable in MG. In section
\ref{sec:numerics}, we present numerical results. First, we investigate
Lane-Emden models of convective stars and calculate the new oscillation periods
and critical value of the adiabatic index. We then turn our attention to
{\scriptsize MESA} models. We calculate the change in the period and inferred
distance from the P-L relation for the same Cepheid models used in
\cite{Jain:2012tn} in order to asses the accuracy of their approximation and
find that when the full effects of hydrodynamic perturbations are included, the
differences from GR can be up to three times larger than the approximation
predicts. Using this, we estimate the new constraints that can be found using
the same data-sets and procedure as \cite{Jain:2012tn}. These correspond to
constraints where the modifications are screened in regions of Newtonian
potential $\Phi_{\rm n}\sim\mathcal{O}(10^{-8})$. Finally, we discuss our
results in light of recent and future astrophysical tests and conclude in
section \ref{sec:concs}.
 
\section{Screened Scalar-Tensor Gravity}\label{sec:mod_grav}

\subsection{The Action}

Any theory which contains a scalar $\phi$ gravitationally-coupled to matter
fields $\psi_{\rm i}$ will generically result in modifications of the Newtonian
force-law in the non-relativistic limit and therefore constitute a modified
theory of gravity. One well-studied class of model has this
coupling arise because the metric $\tilde{g}_{\mu\nu}$ in the matter Lagrangian
differs from the metric appearing in the Ricci scalar $g_{\mu\nu}$ by a
conformal factor $A(\phi)$ such that 
\begin{equation}\label{eq:jfef}
\tilde{g}_{\mu\nu} = A^2(\phi)g_{\mu\nu}.\end{equation} 
Typically, $A(\phi)\approx 1 +\mathcal{O}(\phi/\mpl)$ so that the theories do
not differ too greatly between frames. These theories are described by the
general action
\begin{align}\label{eq:staction}
 S &= \int
\dd^4x\left[\frac{\mpl^2}{2}R-\frac{1}{2}
\nabla_\mu\phi\nabla^\mu\phi-V(\phi)\right]\\&\quad+ S_{\rm m}[\psi_{\rm
i};A^2(\phi)g_{\mu\nu}]
\end{align}
and give rise to a Newtonian fifth-force (per unit mass)
\begin{equation}\label{eq:5force}
 \vec{F}_{\rm \phi} =
\frac{\beta(\phi)}{\mpl}\nabla\phi,\quad\textrm{where}\quad \beta(\phi)\equiv
\mpl\frac{\dd \ln A(\phi)}{\dd \phi}
\end{equation}
is known as the \textit{coupling}. Theories with screening mechanisms generally
have the property that $\dd\beta(\phi)/\dd\phi\approx0$ in the
unscreened region \footnote{Strictly speaking, this is not correct since
$\dd\beta(\phi)/\dd\phi$ is a dimensionful quantity and should be compared
with another physically relevant variable with the same dimensions. In the
unscreened region the field is a small perturbation $\delta\phi$ about
the background field value $\phi_0$ such that $\delta\phi\ll\phi_0$. In this
case one has $(\dd\beta/\dd\phi)\delta\phi\ll\beta(\phi_0)$. When deriving both
the screening mechanism and the perturbation equations in modified gravity
hydrodynamics the terms proportional to $\dd\beta/\dd\phi$ are always suppressed
by factors of $\beta/\delta\phi$ relative to the other terms and so we will
neglect them by setting $\dd\beta/\dd\phi$ to zero.}.
This form of the action is referred to as the \textit{Einstein frame} action
since the \textit{Einstein frame metric}, $g_{\mu\nu}$, is used to compute the
Ricci scalar and not the \textit{Jordan frame metric} $\tilde{g}_{\mu\nu}$. One
may instead work in the \textit{Jordan Frame}, in which case the action can be
obtained using the Weyl rescaling (\ref{eq:jfef}). In what follows we will work
exclusively in the Einstein frame since it allows for intuitive physical
interpretations of the new features of stellar structure and perturbation
theory. A subset of these theories has received a lot of attention recently
because certain choices of the scalar potential $V(\phi)$ and conformal factor
$A(\phi)$ can lead to local screening of the fifth-force (\ref{eq:5force})
through either the chameleon mechanism \cite{Khoury:2003aq,Khoury:2003rn}, the
symmetron mechanism \cite{Hinterbichler:2010es} or the environment-dependent
Damour-Polyakov effect \cite{Brax:2010gi}. In what follows, we will use a
model-independent parametrisation of the theory parameters which is well-suited
for astrophysical tests and refer the reader to \cite{Jain:2010ka,Khoury:2010xi,
Davis:2011qf, Brax:2012gr, Brax:2012yi} for the specific details of
individual screening mechanisms.

Before going on to discuss the high-density screening mechanism, it is important
to address the notion of a conserved matter density. The conformal coupling of
the field to the visible sector has the result that the energy-momentum tensor
of matter $T^{\mu\nu}_{\rm m}=2/{\sqrt{-g}}\delta S_{\rm m}/\delta
g_{\rm\nu}$ is not covariantly conserved, rather one has $\nabla_\mu
T_{\rm m}^{\mu\nu}=\beta(\phi)T_{\rm m} \nabla^{\nu} \phi/\mpl$, where
$T_{\rm m}$ is the trace of the energy-momentum tensor. The density $\rho_{\rm
m} \equiv -T_{\rm m}$ does not obey the usual continuity equation and one
generally works with the conserved density $\rho_{\rm m}\equiv A(\phi)\rho$
instead. This conserved density is also field-independent. From here on we shall
refer to $\rho$ as the conserved matter density corresponding, for example, to
the stellar mass per unit volume.

\subsection{The Screening Mechanism}\label{sec:screenmech}

The coupling to matter has the effect that the scalar potential appearing in the
action (\ref{eq:staction}) is not the same scalar potential which governs the
field's dynamics. Instead, the equation of motion is
\begin{equation}\label{eq:eom}
 \Box\phi = \frac{\dd V(\phi)}{\dd \phi}+\frac{\beta(\phi)\rho}{\mpl}
\end{equation}
corresponding to a density-dependent effective potential 
\begin{equation}\label{eq:veff}
 V_{\rm eff}(\phi) = V(\phi)+\rho A(\phi).
\end{equation}
It is this density-dependence which is responsible for screening the
fifth-force in over-dense environments. Suppose that this potential has a
minimum. The field-position of this minimum can differ by many orders of
magnitude between the solar-system and the cosmological background (recall
$\rho_\oplus/\rho_0\sim 10^{29}$) and so it is possible to engineer effective
potentials where either the mass of small oscillations about the high-density
minimum is large such that the range of the fifth-force is sub-$\mu$m
\cite{Khoury:2003rn} or the coupling $\beta\ll1$ and the magnitude of the force
is negligible \cite{Hinterbichler:2010es,Brax:2010gi}. In this way, the
fifth-force is undetectable in high-density areas.

One can then ask exactly when an object will be screened. Consider a
spherically symmetric over-density with profile $\rho_{\rm b}(r)$ and radius
$R$, which is a small perturbation about a lower but more spatially-extensive
constant density $\rho_{c}$ whose characteristic length-scale is $\gg R$. For
example, a galaxy in the cosmological background or a star inside a galaxy. If
$R$ is large enough such that the field can reach its minimum $\phi_{\rm b}$ at
the higher density and remain there over most of the object's interior then the
fifth-force will be negligible and the object is screened. If the converse is
true and the field deviates only from the value in the background $\phi_{c}$ by
a small perturbation then the full effects of the modification of GR will
manifest. In the static, non-relativistic limit we can ignore time-derivatives
and so the equation of motion (\ref{eq:eom}) is:
\begin{equation}\label{eq:eomnr}
 \nabla^2\phi = V_{,\phi}+\frac{\beta(\phi)\rho}{\mpl}.
\end{equation}

There are then two cases to consider. First, if the field can reach its minimum
we have $V_{,\phi}\approx-\beta\rho/\mpl$ and equation (\ref{eq:eomnr}) is
unsourced. In this case the field is just constant and equal to the
high-density minimum value so that $\phi(r)\approx \phi_{\rm b}$. When this
solution is valid, there are no fifth-forces but in general there is some radius
$\rs$, known as the \textit{screening radius}, where the field begins to
move away from this value and we expect large fifth-forces in this region. In
this second case, the field is a small perturbation about the under-dense value
so that $\phi(r) = \phi_{\rm 0} +\delta\phi(r)$. Subtracting the equation of
motion in both the over and under-dense regions and linearising we have
\begin{equation}\label{eq:screom}
 \nabla^2\delta\phi \approx m_c^2\delta\phi +\frac{\beta_{\rm
c}\delta\rho}{\mpl},
\end{equation}
where $\delta\rho\equiv\rho_{\rm b}-\rho_{c}$, $m_c^2\equiv
V_{,\phi\phi}(\phi_{c})$ is the mass of the field in the background, $\beta_{\rm
c}\equiv\beta(\phi_{\rm c})$ and we have used the fact that $\dd\beta/\dd
\phi\approx 0$. In all theories of interest, we have $m_cR\ll
1$ so that the fifth-force gives rise to novel features on large scales and so
we have ignored the first term, it is negligible compared with the Laplacian.
Now we know that $\delta\rho$ is related to the Newtonian potential $\Phi_{\rm
N}$ via the Poisson equation, $\nabla^2\Phi_{\rm N}=4\pi G\delta\rho$ and so we 
may substitute this into (\ref{eq:screom}) and integrate twice to find the field
profile: 
\begin{widetext}
\begin{equation}
\delta \phi(r) \approx -\phi_{\rm c} + 2\beta_{\rm c} \mpl \left[\Phi_{\rm
N}(r)-\Phi_{\rm
N}(r_{\rm s}) + r_{\rm s}^2\Phi^{\prime}_{\rm N}(r_{\rm s})\left(
\frac{1}{r}-\frac{1}{r_{\rm s}}\right) \right]H(r-r_{\rm s}),
\label{eq:phiform1}
\end{equation}
\end{widetext}
where $H(x)$ is the Heaviside step-function. We refer to the region $r<\rs$ as
the \textit{screened} region and the region $r>\rs$ as \textit{unscreened}.
Using the definition of the Newtonian potential $\dd \Phi_{\rm N}/\dd r
= GM(r)/r^2$, where $M(r)$ is the mass enclosed within a sphere of
radius $r$ (the total mass of the over-density is $M=M(R)$), the fifth-force
(\ref{eq:5force}) in the unscreened region is given by the derivative of
(\ref{eq:phiform1}):
\begin{equation}\label{eq:gprof}
 \frac{\beta_{\rm c}}{\mpl}\frac{\dd\phi}{\dd r} =
\alpha\frac{GM(r)}{r^2}\left[1-\frac{M(\rs)}{M(r)}\right],
\end{equation}
where $\alpha_{\rm c}\equiv 2\beta^2(\phi_{\rm c})$. We will assume that there
are no fifth-forces when $r<\rs$ and that the force is given by equation
(\ref{eq:gprof}) when $r\ge\rs$ from here on\footnote{One may wonder how
accurate this approximation is. \cite{Chang:2010xh} have studied the effects
of chameleon gravity on the structure an evolution of stars by solving the
field equations numerically using a non-linear Gauss-Seidel solver. They have
found that the approximation used here gives results very close to the full
numerical solution and that any differences are negligible.}. When $\rs = R$
there is no fifth-force and the object is screened whereas when $\rs=0$ the
object is full unscreened and we have $F_\phi=\alpha_{\rm c} F_{\rm N}$, where
$F_{\rm N}$ is the Newtonian force. The parameter $\alpha_{\rm c}$ therefore
determines the
strength of the fifth-force. 

All that remains is to find the screening radius $\rs$. Using the fact that
$\delta\phi\rightarrow0$ when $r/\rs\rightarrow\infty$ as does $\phi_{\rm
N}(r)$, equation (\ref{eq:phiform1}) gives us an implicit expression for the
screening radius:
\begin{equation}
\frac{\phi_{\rm c}}{2\beta(\phi_{\rm c}) \mpl}\equiv \chi_{\rm c} = -\Phi_{\rm
N}(r_{\rm s}) -
r_{\rm s}\Phi_{\rm N}^{\prime}(r_{\rm s})\geq0 \label{eq:screenradius}.
\end{equation} 
It will be useful later to recast this as an integral equation:
\begin{equation}\label{eq:chiint}
 \chi_{\rm c} = 4\pi G\int_{\rs}^R r\rho_{\rm b}(r)\dd r.
\end{equation}
The parameter $\chi_{\rm c}$ is known as the \textit{self-screening parameter}
and is of paramount importance to the screening properties of these theories.
Since $\Phi_N <0$ and $\dd\Phi_N/\dd r>0$, there are no solutions when 
\begin{equation}
 |\Phi_{\rm N}(R)|= \frac{GM}{R} > \chi_{\rm c}.
\end{equation}
In this case, the object is fully screened and $\rs=R$. When $\Phi_{\rm
N}>\chi_{\rm c}$ the object will be at least partially unscreened. 

The screening and fifth-force properties in any region are fully specified
by $\alpha_{\rm c}$ and $\chi_{\rm c}$, however
these are very environment-dependent and are non-linearly related to the field
values in different regions of the universe. The exception to this is unscreened
objects, where the field is only a small perturbation around the value in the
background and so these values are roughly constant. Astrophysical tests of MG
\cite{Davis:2011qf,Jain:2011ji,Cabre:2012tq,Jain:2012tn,Vikram:2013uba} are
performed by comparing the properties of screened and unscreened galaxies.
Whether or not a galaxy is screened or not depends on its Newtonian potential
relative to the cosmological value of $\chi_{\rm c}$; the same is true of denser
objects in unscreened galaxies. In each case, the strength of the fifth-force in
any unscreened region is proportional to the cosmological value of $\alpha_{\rm
c}$ and so astrophysical tests of MG probe the cosmological values of
$\alpha_{\rm c}$ and $\chi_{\rm c}$. The theory is then parametrised in a
model-independent manner by the cosmological values of these parameters, which
we denote by $\alpha$, which measures the strength of the fifth-force in
unscreened regions and $\chi_0$, which determines how small the Newtonian
potential must be in order for the object to be unscreened\footnote{One must
note however that this analysis applies to isolated objects only. In the
presence of other objects of Newtonian potential $\Phi_{\rm N}^{\rm ext}$ the
external potential can be large enough to screen an object that would otherwise
be self-unscreened \cite{Hui:2009kc}. This environment dependence has formed the
basis for many observational tests of these theories
\cite{Davis:2011qf,Jain:2011ji,Cabre:2012tq,Jain:2012tn,Vikram:2013uba}
and indeed the new effects presented here will rely on this feature as a
potential probe.}. We will work with this parametrisation in what follows.
$f(R)$ theories are chameleon theories with $\alpha=1/3$ \cite{Brax:2008hh} and
$\chi_0$ is equivalent to the derivative of the function today, $f_{R0}$.

\subsection{Astrophysical Tests of Modified Gravity}

We will not attempt any comparison with data here but for completeness and
motivation we will describe the basic premise behind observational tests using
astrophysical effects below.

\subsubsection{Current Constraints}

Only unscreened objects will show deviations from GR and so we must look for
such objects in the universe if we wish to observe these deviations. There are
two classes of screened objects: self-screened, where the self-Newtonian
potential of the object is larger than $\chi_0$ and environment-screened, where
the object would be unscreened in isolation but is still screened due to the
Newtonian potential of its neighbours (see \cite{Hui:2009kc} for a nice
discussion of equivalence principle violation effects and their relation to
environmental screening). The cosmological bound coming from cluster
constraints is $\chi_0\lsim 10^{-5}$ \cite{Schmidt:2008tn}. The Newtonian
potential of spiral galaxies is $\mathcal{O}(10^{-6})$ and so this would leave
most galaxies, including our own milky way self-unscreened. Clearly if this
were the case then our own galaxy would show deviations from GR, which are not
observed and so we must conclude that our own galaxy is screened, either
because $\chi_0\lsim10^{-6}$ or because we are screened by the Newtonian
potential of the local group. While this has been debated in the past,
recently, an independent constraint from comparing water maser distances to NGC
4258 with tip of the red giant branch distances has placed the bound
$\chi_0<10^{-6}$ \cite{Jain:2012tn}. Finally, the strongest constraints on
$\chi_0$ come from comparing tip of the red giant branch distances with Cepheid
distances and place the bound $\chi_0\lsim4\times10^{-7}$ \cite{Jain:2012tn}. 

\subsubsection{Dwarf Galaxy Tests}

With this bound, there are very few unscreened objects in the universe. There
are no cosmological signatures, either at the level of the background or linear
perturbations (although see \cite{Brax:2012mq,Brax:2013yja} for an
unusual model with a very strong coupling to matter) and spiral galaxies
are screened. The only self-unscreened objects in the universe are dwarf
galaxies, whose Newtonian potentials are $\mathcal{O}(10^{-8})$ and large
post-main-sequence stars \footnote{HI gas clouds in galaxies have Newtonian
potentials of order $10^{-11}$ \cite{Hui:2009kc}, although, to-date, no
observational tests exploiting this have been attempted.}. Isolated dwarf
galaxies located in voids should therefore show deviations from GR compared to
their counterparts located in clusters. The basic premise then is to compare the
observational properties altered by MG in a sample of screened and unscreened
dwarf galaxies. Any statistically significant discrepancy is a sign of MG
whereas an agreement within experimental errors constrains $\chi_0$ and
$\alpha$. One then needs some method of discerning whether or not an observed
galaxy is screened. Recently, the results of MG N-body simulations have been
used in conjunction with Sloan Digital Sky Survey (SDSS) data to provide a
screening map of the universe
for different values of $\chi_0$ \cite{Cabre:2012tq}. Using this map, it has
already been possible to place new constraints on these theories
\cite{Jain:2012tn,Vikram:2013uba}.

Observational signatures using dwarf galaxies generally fall into two classes:
modifications of the galaxy dynamics \cite{Jain:2011ji} and those resulting
from effects on stellar structure \cite{Chang:2010xh,Davis:2011qf}. Both have
been carried out \cite{Jain:2012tn,Vikram:2013uba} using SDSS data and,
currently, the former is not competitive with the latter. In this work we are
interested in stellar structure effects and so we shall only describe these
here.

The most practical method for using stellar structure tests is to compare
different distance indicators to the same galaxy. The distance to a
galaxy using stellar phenomena is either calibrated on Milky Way (or local
group galaxies in some cases) objects or derived from the GR theoretical model.
Either way, if the stellar property used to calculate this distance is altered
in MG then the measured distance to an unscreened galaxy will differ from the
GR distance. If, on the other hand, the indicator is insensitive to MG then
any observation will give precisely the GR distance. If one then compared two
distances to the same galaxy, one using a screened distance indicator and the
other using an unscreened one, any discrepancy probes MG. By using many
different galaxies and comparing to a screened sample, a statistically
significant constraint can be achieved. This is what was done in
\cite{Jain:2012tn}. 

\subsubsection{Cepheid Distances in Modified Gravity}

In practice, the best unscreened distance indicators are Cepheid variable stars.
They have Newtonian potentials of $\mathcal{O}(10^{-7})-\mathcal{O}(10^{-8})$
and there is a large amount of data available.

Cepheid stars are 5-10 $M_\odot$ stars which have evolved off the main-sequence
and onto the red giant branch. During this phase, their Hertzsprung-Russell (HR)
diagrams show a so-called blue loop, where the temperature rapidly increases
and decreases at approximately constant luminosity. Whilst traversing the
loop, the tracks cross the \textit{instability strip}, where the star is
unstable to \textit{Cepheid pulsations}. Along this narrow strip in the $\log
L$-$\log T_{\rm eff}$ plane, the star is unstable due to a layer of partially
ionised helium coinciding with the region where the motion becomes
non-adiabatic. Upon compression, the energy does not go into raising the
pressure, which would result in an increased outward force giving a stabilising
effect but rather into further ionising the helium. This damming up of the
energy acts like an engine and drives large changes, up to 10\%, in the
stellar radius. Along this strip, there is a well-known period luminosity
relation:
\begin{equation}
 M_{\rm B} = a\log\tau + b\log({\rm B-V})+c,
\end{equation}
where $M_{\rm B}$ is the bolometric magnitude and $a$, $b$, and $c$ are
empirically determined constants. Empirically, $a\approx-3$
\cite{Freedman:2010xv}. Now in MG, and we will see this derived below, the
period $\tau$ of a Cepheid at fixed luminosity, effective temperature and mass
change by some amount $\Delta \tau = \tau_{\rm MG}-\tau_{\rm GR}$ so that this
empirically-fitted relation predicts a deviation from the GR distance
\begin{equation}\label{eq:plmg}
 \frac{\Delta d}{d} = -\frac{1}{2}a\frac{\Delta \tau}{\tau}. 
\end{equation}
Previously \cite{Jain:2012tn}, $\Delta \tau$ was calculated using the
approximation $\Delta\tau/\tau\approx-\Delta G/2G$, where $\Delta G$ is
calculated using some appropriate average. In this work, we will derive the
equations governing its value and show how it can be calculated numerically
using some simple examples and compare our results with this approximation.  

\section{Modified Gravity Hydrodynamics}\label{sec:MGhydro}

In this section we shall start from the modified equations of hydrodynamics and
proceed to derive the equations governing both the equilibrium structure and
linearised radial perturbations of non-relativistic stars.

We will describe bulk quantities such as the pressure and density in the
\textit{Eulerian picture}, where these quantities are to be considered as fields
which give the value of said quantities at any point in space as a function
of time. In contrast, we will describe the position of individual fluid
elements (and, when needed, the pressure perturbations) in the
\textit{Lagrangian picture}, where the motion of individual fluid elements are
followed as a function of time. In this case, the Lagrangian position $\vec{r}$
of a fluid element satisfies the momentum equation:
\begin{equation}\label{eq:momentumgen}
 \frac{\partial^2 \vec{r}}{\partial t^2} = -\frac{1}{\rho}\nabla P + \vec{F},
\end{equation}
where $P(\vec{r})$ is the pressure, $\rho(\ver)$ is the density and $\vec{F}$
is the external force density. In MG, the fluid moves under its own Newtonian
gravity and the fifth-force
(\ref{eq:5force}) due to the scalar field so that
\begin{equation}\label{eq:momentumMG}
 \frac{\partial^2 \vec{r}}{\partial t^2} = -\frac{1}{\rho}\nabla P -
\frac{GM(r)}{r^3}\ver-\frac{\beta(\phi)}{\mpl}\nabla\phi.
\end{equation}
This is the only hydrodynamical equation that is altered relative to GR;
changing gravity only changes the motion of the fluid elements and does not
directly\footnote{There may be indirect changes due to the change in the
structure of the fluid, for example the surface temperature, but these are not
the direct result of a gravitational interaction but rather a response of the
system to a change in the force \cite{Davis:2011qf}.} alter other processes such
as mass conservation, energy generation and radiation flux. The quantity $M(r)$
($r\equiv|\ver|$) is the mass enclosed inside a radius $r$ from the centre, and
is given via the Poisson equation
\begin{equation}\label{eq:poisson}
 \nabla^2\Phi_{\rm N}=4\pi G\rho(r),
\end{equation}
which may be integrated once to give 
\begin{equation}\label{eq:poissint}
 \frac{\dd \pn}{\dd r}=\frac{GM(r)}{r^2}.
\end{equation} 
Since mass is a locally-conserved quantity we also have the continuity
equation:
\begin{equation}\label{eq:cont}
 \frac{\partial \rho}{\partial t}+\vec{\nabla}\cdot(\rho\vec{v})=0,
\end{equation}
where $\vec{v}\equiv\dd \ver/\dd t$ is the velocity of the fluid element. In
general, one must also consider the energy generation and radiative transfer
equations but these are only important if one wishes to study the effects
of perturbations coupled to stellar atmospheres, which is irrelevant in the
context of modified gravity. We will include their effects when simulating the
equilibrium stellar configuration numerically in order to produce the correct
stellar properties, however, we will not include them in our perturbation
analysis. Instead, we will work in the so-called \textit{adiabatic}
approximation, where the density and pressure evolve according to
\begin{equation}\label{eq:dpdrho}
 \frac{\dd P}{\dd t} = \frac{\Gamma_1 P}{\rho}\frac{\dd \rho}{\dd t}.
\end{equation}
The quantity
\begin{equation}\label{eq:Gamma1def}
 \Gamma_1 \equiv \left(\frac{\dd \ln P}{\dd \ln \rho}\right)_{\rm adiabatic}
\end{equation}
is the \textit{first adiabatic index} or \textit{equation of state}. It is of
paramount importance to the study of stellar pulsation and stability and we
will return to discuss it later on. The simplest gases are typically
characterised by a constant equation of state so that
$P\propto\rho^{\Gamma_1}$, where $\Gamma_1 = 5/3$ ($4/3$) for non-relativistic
(relativistic) gases. In general, one must couple the hydrodynamic equations to
the full radiative transfer system and extract the equation of state from the
result. 

\subsection{Equilibrium Structure}\label{sec:eqstruc}

We are interested in small perturbations about the equilibrium structure of
stars and so we must first find the modified zeroth-order equations. These were
presented heuristically in \cite{Davis:2011qf}, however, here we shall
derive them from this more formal set-up. The equilibrium stellar configuration
is both static and spherically symmetric and can be found by setting
time-derivatives to zero and $\ver= r$ in the hydrodynamic equations so that
this now represents the Eulerian coordinate. We will denote all equilibrium
quantities with a subscript-zero except for $M(r)$, which is defined at the
background level only. It is important to note that $\chi_0$ is not a property
of the star but is the cosmological value of $\chi_{\rm c}$ found by evaluating
(\ref{eq:screenradius}) using the cosmological values of $\phi$ and $\beta$. In
what follows, $\phi_0(r)$ is the equilibrium field-profile throughout the star
and not the cosmological value. With no time-dependence, (\ref{eq:cont}) is
trivially satisfied and $\rho(r,t)=\rho(r)$. This simple form of the
density profile allows us to find the mass enclosed in any given radius:
\begin{equation}\label{eq:masscons}
 \frac{\dd M(r)}{\dd r}=4\pi r^2\rho_0(r).
\end{equation}
The momentum equation (\ref{eq:momentumMG}) then reduces to the \textit{modified
hydrostatic equilibrium equation} 
\begin{equation}\label{eq:MGHSE}
 \frac{\dd P_0(r)}{\dd r} =
-\frac{GM(r)\rho_0(r)}{r^2}-\frac{\beta(\phi_0)\rho_0(r)}{\mpl}\frac{\dd
\phi_0(r)}{\dd r}.
\end{equation}
Physically, this equation describes the pressure profile the star must assume
in order for the star to support itself against gravitational collapse. The
second term is the fifth-force due to the scalar field; stars in modified
gravity need to provide larger pressure gradients in order to combat this extra
inward component \cite{Chang:2010xh,Davis:2011qf,Jain:2012tn}. These equations
are then supplemented by the radiative transfer equation
\begin{equation}\label{eq:radtrans}
 \frac{\dd T_0(r)}{\dd r} = -\frac{3}{4 a} \frac{\kappa(r)}{T_0^3}
\frac{\rho_0(r) L_0(r)}{4\pi r^2},
\end{equation}
which describes how the temperature $T(r)$ varies due to the flux of energy with
luminosity $L(r)$ away from regions of energy generation governed by
\begin{equation}\label{eq:engen}
 \frac{\dd L_0}{\dd r} = 4\pi r^2 \rho_0(r) \epsilon(r).
\end{equation}
Here, $\kappa(r)$ is the \textit{opacity}, the cross-section for radiation
absorption per unit mass and $\epsilon(r)$ is the energy generation rate per
unit mass from processes such as nuclear burning.

Taken by themselves, these equations do not close and one must specify the
equations of state relating $P_0, \rho_0, \kappa$ and $\epsilon$, which are
themselves determined by further equations involving energy transfer and nuclear
burning networks. In the subsequent sections, we will solve these equations
analytically by assuming different equations of state. When computing the
equilibrium structure using {\scriptsize MESA}, look-up tables are used to find
the opacity and pressure at a given density and temperature while energy
generation rates are found using nuclear burning networks for individual
elements. We refer to reader to the {\scriptsize MESA} instrumentation papers
\cite{Paxton:2010ji,Paxton:2013pj} for the full details.

\subsection{Linear Perturbations}

The dynamics of stellar oscillations are governed by the hydrodynamics of small
perturbations about the equilibrium configuration and so we shall linearise the
equations (\ref{eq:momentumMG}), (\ref{eq:poisson}), (\ref{eq:cont}) about some
assumed background profile $\{P_0(r), \rho_0(r), \pno(r), \gao(r), \phi_0(r)\}$
ignoring second-order terms in the perturbations. 

The fundamental object of interest is the linearised perturbation to the
Lagrangian position of each fluid element $\vec{\delta r}(\vec{r})$, which
describes the oscillation of the fluid from equilibrium at each radius. So far,
our treatment has been completely general and we have worked with the full
three-dimensional problem. Continuing in this fashion would result in a
complete treatment of both radial and non-radial modes of oscillation. The aim
of this work is to provide a consistent framework with which to predict the
oscillation properties of partially unscreened stars residing in unscreened
dwarf galaxies. The extra-galactic nature of these stars ensures that only their
fundamental radial mode (and possibly the first-overtone) are
observable and so from here on we will specify to the case of radial
oscillations so that $\vec{\delta r}(r)$ is a purely radial vector\footnote{One
could derive the full, fourth-order system of equations governing non-radial
oscillations, however, its solutions are irrelevant for observational tests of
modified gravity.}. A more convenient quantity to work with is the relative
displacement, given by
\begin{equation}\label{eq:zetadef}
 \zeta(r) \equiv \frac{\delta r(r)}{r},
\end{equation}
and the goal of the present section is to derive is equation of motion.

We shall work with Eulerian perturbations of the background profile, which we
will distinguish from Lagrangian perturbations by the use of a tilde so that
\begin{align}\label{eq:backgroundperts}
 P(r,t)&=P_0(r)+\tilde{P}(r,t)\\
\rho(r,t)&=\rho_0(r)+\tilde{\rho}(r,t)\\
\Phi_{\rm N}(r)&=\Phi_{{\rm N}, 0}(r)+\tilde{\Phi}_{\rm N}(r,t)\\
\phi(r,t)&=\phi_0(r)+\tilde{\phi}(r,t).
\end{align}
As remarked above, the Lagrangian perturbations may provide more physical
insight on occasion and the two are related, for example, via
\begin{align}\label{eq:perteg}
 \delta P(r,t) &= P(\vec{r}+\vec{\delta r},t)-P_0(r) \\&=
\tilde{P}(r,t)+\delta r\frac{\dd
P_0(r)}{\dd r}.
\end{align}
It will be useful later to have the Lagrangian
pressure perturbation in terms of our system-variables. This is given by
\begin{equation}\label{eq:lagpresspert}
 \frac{\delta P}{P_0}= \gao\frac{\delta\rho}{\rho_0} =
-\gao\left(3\zeta+r\frac{\dd \zeta}{\dd r}\right),
\end{equation}
where the first equality holds only in the adiabatic approximation.

We begin by perturbing equation (\ref{eq:cont}) to find
\begin{equation}\label{eq:contpert}
 \tilde{\rho}=-\frac{1}{r^2}\frac{\partial }{\partial r}(r^3\rho_0\zeta).
\end{equation}
This may be used in the perturbed form of (\ref{eq:momentumMG}) to
find\footnote{The
sharp-eyed reader will notice that there is no term proportional to
$\dd\beta/\dd\phi$ as one would expect. Technically, such a term should be
present, however, we have already seen in section (\ref{sec:screenmech}) that
this is negligible in theories of interest. In fact, it is a requirement of the
screening mechanism that such derivatives are negligible and so we have set
this term to zero when perturbing the momentum equation.}
\begin{equation}\label{eq:accpert}
 \rho_0r\frac{\partial^2\zeta}{\partial t^2}=-\frac{\dd \tilde{P}}{\dd
r}-\rho_0\frac{\dd\tilde{\Phi}_{\rm N}}{\dd r}+\frac{\beta_0}{\mpl
r^2}\frac{\partial }{\partial r}(r^3\rho_0\zeta)\frac{\dd \phi_0}{\dd
r}-\frac{\beta_0\rho_0}{\mpl}\frac{\dd \tilde{\phi}}{\dd r}.
\end{equation}
The perturbed Poisson equation is
\begin{equation}\label{eq:pertpoisson}
 \nabla^2\tilde{\Phi}_{\rm N}=4\pi G\tilde{\rho},
\end{equation}
which may be integrated once using (\ref{eq:contpert}) to find
\begin{equation}
 \frac{\dd \tilde{\Phi}_{\rm N}}{\dd r}=-4\pi Gr\rho_0\zeta.
\end{equation}
Now 
\begin{equation}
 \frac{\dd P}{\dd t}=\frac{\partial P}{\partial t}+\vec{v}\cdot\nabla
P=\frac{\partial \tilde{P}}{\partial t}+r\frac{\partial \zeta}{\partial
t}\frac{\partial P_0}{\partial r}
\end{equation}
and using (\ref{eq:dpdrho}) this is 
\begin{equation}\label{eq:pertprho}
 \tilde{P}+r\zeta\frac{\partial P_0}{\partial
r}=\frac{\Gamma_{1,0}P_0}{\rho_0}\left(\tilde{\rho}+r\zeta\frac{\partial
\rho_0}{\partial r}\right).
\end{equation}
We wish to eliminate $\tilde{P}$ and $\tilde{\rho}$ and so we substitute
(\ref{eq:contpert}) into (\ref{eq:pertprho}) to find
\begin{align}
 \tilde{P}&=-\frac{\Gamma_{1,0}P_0}{\rho_0}\left(\frac{1}{r^2}\frac{\partial}{
\partial r}(r^3\rho_0\zeta)-r\zeta\frac{\dd P_0}{\dd
r}+\frac{\rho_0}{\Gamma_{1,0}P_0}\frac{\dd P_0}{\dd r}r\zeta\right)\nonumber\\
&=-\Gamma_{1,0}P_0r\left(\frac{\partial \zeta}{\partial
r}+\frac{3}{r}\zeta+\frac{1}{\Gamma_{1,0}P_0}\frac{\dd P_0}{\dd r}\zeta\right).
\end{align}
We then have 
\begin{align}
 \frac{\partial \tilde{P}}{\dd r}&=-\frac{\partial}{\partial
r}(\Gamma_{1,0}rP_0\frac{\partial\zeta}{\partial
r})\nonumber\\&-\frac{\partial}{\partial
r}(3\Gamma_{1,0}P_0\zeta)-\frac{\partial}{\partial r}(r\frac{\dd P_0}{\dd
r}\zeta).
\end{align}
\newline\newline 
This may then be used with (\ref{eq:pertpoisson}) and
(\ref{eq:accpert}) to
find:
\begin{align}
 r^4\rho_0\frac{\partial ^2 \zeta}{\partial t^2}&=\frac{\partial}{\partial
r}\left(r^4\Gamma_{1,0}P_0\frac{\partial \zeta}{\partial
r}\right)+r^3\frac{\partial}{\partial
r}\left[\left(3\Gamma_{1,0}-4\right)P_0\right]\zeta\nonumber\\&-\frac{
r^3\beta_0\rho_0}{ \mpl } \frac{\partial \tilde{\phi}}{\partial
r}-\frac{\beta_0\rho_0}{\mpl}r^4\frac{\dd^2\phi_0}{\dd
r^2}\zeta-2\frac{r^3\beta_0\rho_0}{\mpl}\frac{\dd \phi_0}{\dd r}\zeta.
\end{align}
This is the equation of motion governing the evolution of $\zeta$. In GR,
$\partial \tilde{\phi}/\partial r=0$ and we obtain a single equation. In MG,
however, we need a separate equation for $\tilde{\phi}$, which is found by
perturbing the equation of motion (\ref{eq:eom}):
\begin{equation}
 -\frac{\partial^2\tilde{\phi}}{\partial
t^2}+\nabla^2\tilde{\phi}=m_0^2\tilde{\phi}-3\frac{\beta}{\mpl}
\rho_0\zeta-\frac{\beta}{\mpl}r\frac{\partial }{\partial r}(\rho_0\zeta),
\end{equation}
where $m_0^2\equiv V_{\phi\phi}(\phi_0)$ is the mass of the unperturbed
field  at zero density and equation (\ref{eq:contpert}) has been used. We are
interested in stationary-wave solutions and so we expand
\begin{align}
 \zeta(r,t)&=\xi(r)e^{i\omega t}\\
\tilde{\phi}(r,t)&=\varphi(r)e^{i \omega t}
\end{align}
to yield two coupled equations
\begin{widetext}
\begin{align}
(\nabla^2+\omega^2)\varphi& = m_0^2\varphi-3\frac{\beta}{\mpl}\rho_0\xi-\frac{
\beta}{\mpl}r\frac{\partial }{\partial r}(\rho_0\xi),\label{eq:varphi}\\
\frac{\dd}{\dd r}\left(r^4\Gamma_{1,0}P_0\frac{\dd \xi}{\dd
r}\right)+r^3\frac{\dd}{\dd
r}\left[\left(3\Gamma_{1,0}-4\right)P_0\right]\xi&-\frac{r^4\beta_0\rho_0}{\mpl}
\frac{\dd^2\phi_0}{\dd r^2}\xi-2\frac{r^3\beta_0\rho_0}{\mpl}\frac{\dd
\phi_0}{\dd r}\xi+r^4\rho_0\omega^2\xi=\frac{r^3\beta_0\rho_0}{\mpl}
\frac { \dd \varphi}{\dd r}.\label{eq:MLAWE}
\end{align}
\end{widetext}
Equations (\ref{eq:varphi}) and (\ref{eq:MLAWE}) constitute the main result of
this
section. One could combine them into a single equation, however, it is more
instructive to treat the system as two coupled equations. In GR, we have
$\varphi=\beta_0=\dd\phi_0/\dd r = 0$ and (\ref{eq:MLAWE})
reduces to
\begin{align}\label{eq:LAWE}
\frac{\dd}{\dd r}&\left(r^4\Gamma_{1,0}P_0\frac{\dd \xi}{\dd
r}\right)\nonumber\\&+r^3\frac{\dd}{\dd
r}\left[\left(3\Gamma_{1,0}-4\right)P_0\right]\xi+r^4\rho_0\omega^2\xi=0,
\end{align}
which describes linear, adiabatic, radial waves moving in the stellar interior.
It is known at the \textit{linear adiabatic wave equation} (LAWE) and its
eigenfrequencies $\omega$ give the frequency of stellar oscillations about
equilibrium. We will hence refer to (\ref{eq:MLAWE}) as the \textit{modified
linear adiabatic wave equation} (MLAWE). Its properties will be the subject of
the next section.

Using the profile (\ref{eq:gprof}), we have
\begin{align}\label{eq:mgcham}
\frac{\beta_0}{\mpl}\left[2\frac{\dd\phi_0}{\dd r}+r\frac{\dd^2\phi_0}{\dd
r^2}\right]=4\pi\alpha Gr\rho_0(r)\quad r>\rs,
\end{align}
which we shall use in all analytic and numerical computations from here on.

\subsection{Application to Other Modified Gravity Theories}

Before proceeding, it is worth noting that although this work focuses on
chameleon-like theories, the hydrodynamic analysis performed so far is more
general than this. Any scalar-tensor theory whose Einstein frame matter coupling
can be written in the effective form
\begin{equation}
 \mathcal{L}_{\rm coupling}=C(\phi)T_{\rm m},
\end{equation}
where $T_{\rm m}$ is the trace of the matter energy-momentum tensor, will give
rise to an additional force
\begin{equation}
 \vec{F}_\phi=\frac{\beta(\phi)}{\mpl}\nabla\phi\quad\beta(\phi)\equiv
\mpl\frac{\dd C(\phi)}{\dd \phi},
\end{equation}
provided that $\phi$ is the canonically normalised field \footnote{If this is
not the case one can either canonically normalise if possible or define a more
general form of $\beta(\phi)$ in terms of $C(\phi)$ and the field-dependent
factor multiplying the kinetic term.}. The only assumption we have made so far
is $\dd\beta(\phi)/\dd\phi\approx0$ but this was not necessary, only
appropriate, and if one has a scalar-tensor theory where this is not the case
then one would find additional terms in (\ref{eq:MLAWE}). In terms of the
theories studied so far we have $C(\phi)=\ln A(\phi)$. Another class of theories
that has attracted recent interest are those which screen via the Vainshtein
mechanism \cite{Vainshtein:1972sx} such as Galileons \cite{Nicolis:2008in} and
massive gravity \cite{deRham:2010kj,Hinterbichler:2011tt}, which are described
by $C(\phi)=\gamma\phi/\mpl$ where $\gamma$ is a constant describing the
strength of the fifth-force. In this case, one has $\beta(\phi)=\gamma$ so that
$\dd\beta/\dd\phi = 0$. The MLAWE (\ref{eq:MLAWE}) therefore applies equally to
Vainshtein screened theories provided that one provides the counterpart to the
equation of motion for the field perturbation (\ref{eq:varphi}). Vainshtein
screening is too efficient to show any modified gravity effects on stellar
structure at the background level and so one would expect the terms
proportional to derivatives of $\phi_0$ to be negligible and hence that the
oscillation frequencies are identical to GR, however, this is not necessarily
true at the level of perturbations and scalar radiation may provide an
observational test. We will not investigate this possibility here but postpone
it for future work. 

\section{Properties of the Modified Linear Adiabatic Wave
Equation}\label{sec:MLAWE}

The MLAWE describes the behaviour of stellar oscillations under modified
gravity. There are two major differences with respect to the GR equation.
Firstly, there are two additional terms in the homogeneous part, proportional to
the first and second derivatives of the background field. When $r<\rs$ these
are negligible and the homogeneous part behaves as it would in GR,
however, these are comparable to the other terms when $r>\rs$ and
encode the effect of modified gravity on wave propagation in the region
exterior to the screening radius. Physically, the term proportional to $\dd
\phi_0/\dd r$ acts as a varying enhancement of Newton's constant $G(r)$ given
by equation (\ref{eq:gprof}) and the term proportional to $\dd^2
\phi_0/\dd r^2$ can schematically be viewed as $\dd G(r)/\dd r$ and so it
encodes the effect of a varying Newtonian force in the outer regions.

The second effect is a driving term proportional to $\beta_0\dd \varphi/\dd r$.
This is clearly the effect of the fifth-force due to perturbations in the
field. This was modeled by \cite{Silvestri:2011ch} as an inhomogeneous forcing
term at a single frequency. Here it appears as coupling between the field and
stellar perturbations: the stellar perturbations source the scalar field
perturbations and vica versa. Physically, $\tilde{\phi}(r,t)$ corresponds to
scalar radiation (or rather the flux $T_{\phi\,0i}$ at infinity). There is
evidence from previous studies \cite{Silvestri:2011ch,Upadhye:2013nfa} that
this is negligible in chameleon-like systems and so from here on we will neglect
the dynamics of the field perturbations and treat only the homogeneous part of
the MLAWE (\ref{eq:MLAWE}). There may, nevertheless, be interesting features
associated with the coupled oscillator-like problem such as beating and scalar
radiation and this is left for future work.

\subsection{Boundary Conditions}\label{sec:BCS}

The MLAWE requires two boundary conditions in order to fully specify the
solution given a specific value of $\omega$\footnote{Of course, we also need to
derive the value of $\omega^2$ from the solution of the equation. This can be
done by looking for values such that the solutions satisfy both boundary
conditions and is discussed in section (\ref{sec:evalue}).}. Firstly, our system
is spherically symmetric and so we must impose $\delta r=0$ at $r=0$. The MLAWE
then requires
\begin{equation}\label{eq:BCC}
 \left.\frac{\dd \xi}{\dd r}\right\vert_{r=0}=0.
\end{equation}
The surface boundary condition depends on the stellar atmosphere model (see
\cite{cox1980theory} for a discussion) but the lowest modes, where the
period of oscillation is far longer than the inertial response time of the
atmosphere can be described by solutions with $P(R)=0$ so that $\delta P(R)=0$.
This gives the surface condition \cite{cox1980theory}
\begin{equation}\label{eq:BCS}
 \left.\frac{\delta P}{P_0}\right\vert_{r =
0}=\left(\frac{\omega^2R^3}{GM}+4\right)\zeta(R),
\end{equation}
where the Lagrangian pressure perturbation is given by (\ref{eq:lagpresspert}).
Note that the additional terms in the MLAWE vanish at the stellar centre and
radius if we take $\rho_0(R)=0$ so that these conditions are identical to GR.

\subsection{Sturm-Liouville Nature of the Equation}\label{sec:evalue}

The MLAWE can be written in Sturm-Liouville form
\begin{equation}
 \hat{\mathcal{L}}\xi + w(r)\omega^2\xi,=0
\end{equation}
where the weight function $w(r)=r^4\rho_0(r)$ and the operator can be written
\begin{align}
 \hat{\mathcal{L}}^{\rm GR}&=\frac{\dd}{\dd
r}\left(r^4\Gamma_{1,0}P_0\frac{\dd}{\dd r}\right)+r^3\frac{\dd}{\dd
r}\left[\left(3\Gamma_{1,0}-4\right)P_0\right],\\
\hat{\mathcal{L}}^{\rm MG}&=
\hat{\mathcal{L}}^{\rm GR}-\frac{\beta\rho_0}{\mpl}r\frac
{\dd^2\phi_0}{\dd r^2}-2\frac{\beta\rho_0}{\mpl}\frac{\dd \phi_0}{\dd r}.
\end{align}
The problem of finding the pulsation frequencies is then one of finding the
eigenvalues of these equations that correspond to eigenfunctions satisfying the
boundary conditions (\ref{eq:BCC}) and (\ref{eq:BCS}). In practice, it is not
possible to solve these equations analytically for physically realistic stars
and numerical methods must be used. We will do just this in section
\ref{sec:numerics}. Despite the need for numerics, a lot of the new modified
gravity features can be discerned and elucidated using well-known
Sturm-Liouville techniques and so we shall investigate these first. 

\subsection{Scaling Behaviour of the Eigenfrequencies}\label{sec:scaling}

Using the dimensionless quantities:
\begin{align}
 \bar{P}_0(r)&\equiv\frac{R^4}{GM^2}P_0(r),\label{eq:dimlessP}\\ 
\bar{\rho}_0(r)&\equiv\frac{R^3}{M}\rho_0(r)\quad\textrm{and}
\label{eq:dimlessrho} \\ x&\equiv
\frac{r}{R},\label{eq:dimlessr}
\end{align}
the MLAWE (\ref{eq:MLAWE}) can be cast into dimensionless form:
\begin{align}\label{eq:dimlessLAWE}
 &\frac{\dd}{\dd x}\left(x^4\gao\bar{P}_0\frac{\dd \xi}{\dd
x}\right)\\&+x^3\frac{\dd}{\dd
x}\left[\left(3\gao-4\right)\bar{P}_0\right]\xi+x^4\bar{\rho}_0\left[
\Omega^2-4\pi\alpha\bar{\rho}_0\right]=0,
\end{align}
where
\begin{equation}\label{eq:dimOmega}
 \Omega^2\equiv\frac{\omega^2R^3}{GM}
\end{equation}
is the dimensionless eigenfrequency and the term proportional to $\alpha$ is
only present when $r>\rs$. In GR, $\alpha=0$ and one can solve this given some
equilibrium stellar model to find $\Omega^2$. Since this must be a
dimensionless number one has $\omega^2\propto GM/R^3$. In MG, there are two
sources of change to this value at fixed mass: the change due to the different
equilibrium structure described by (\ref{eq:MGHSE}) and the change due to the
additional term in the MLAWE. At the level of the background,
we expect that stars of fixed mass have smaller radii and larger values of
$\langle G\rangle$ (where by $\langle\rangle$ we mean some appropriate average
over the entire star) so that the frequencies are higher in MG and at the
level of perturbations, one can replace $\Omega^2$ in the GR equation by the
effective frequency $\Omega^2-4\pi\alpha\langle\bar{\rho}_0\rangle$ so that
$\Omega^2_{\rm MG}\approx \Omega^2_{\rm
GR}+4\pi\alpha\langle\bar{\rho}_0\rangle$ and we therefore expect the MG
eigenfrequency to be larger still.

The behaviour and importance of each effect requires a full numerical treatment,
however, one can gain some insight by considering scaling relations when the
star is fully unscreened so that $G\rightarrow G(1+\alpha)$. The equations of
stellar structure are self-similar so that we can replace each quantity, such
as the pressure, by some characteristic quantity, in this example the central
pressure $P_{\rm c}$, multiplied by some dimensionless function in order to see
how these different characteristic quantities scale when others are varied. Let
us assume an equation of state of the form 
\begin{equation}\label{eq:eosgam}
P=K\rho^\gamma,
\end{equation}
where $K$ is a
constant and $\gamma$ differs from $\gao$ since the system need not be
adiabatic. In this case, equations (\ref{eq:masscons}) and (\ref{eq:MGHSE})
give
\begin{align}
 \rho_{\rm c}&\propto \frac{M}{R^3}\\
\rho_{\rm c}^{\gamma-1}&\propto \frac{GM^2}{R},
\end{align}
which can be combined to find the scaling of the radius for a fully-unscreened
star in MG:
\begin{equation}\label{eq:Rscaling}
 \frac{R_{\rm MG}}{R_{\rm GR}}=(1+\alpha)^{-\frac{1}{3\gamma-4}}
\end{equation}
at fixed mass. Ignoring the MG perturbations, one would
then expect
\begin{equation}\label{eq:backgroundomegascaling}
 \frac{\omega^2_{\rm MG}}{\omega^2_{\rm
GR}}=(1+\alpha)^{\frac{3\gamma-1}{3\gamma-4}}.
\end{equation}
We shall confirm this limit numerically for some simple models later in section
\ref{sec:LEmodels}. In the fully unscreened limit, we would then expect the
eigenfrequencies to scale approximately like
\begin{equation}
 \frac{\omega^2_{\rm MG}}{\omega^2_{\rm
GR}}\sim(1+\alpha)^{\frac{3\gamma-1}{3\gamma-4}}\left(1+
\frac{4\pi\alpha}{\Omega_{\rm GR}^2}\langle\bar
{ \rho }_0\rangle\right)
\end{equation}
so that they are always larger than the GR prediction, at least when
$\omega^2_{\rm GR}>0$.

\section{Stellar Stability}\label{sec:stellarstability}

Given the Sturm-Liouville nature of the problem, we can find an upper bound on
the fundamental frequency $\omega_0$ using the variational principle. Given an
arbitrary trial function $\Psi(r)$, one can construct the functional
\begin{equation}
 F[\omega]\equiv-\frac{\int_0^R\dd r \,\Psi^*(r)\hat{\mathcal{L}}\Psi(r)
}{\int_0^R\dd r\,\Psi^*(r)\Psi(r)\rho_0r^4},
\end{equation}
which has the property that $\omega_0^2\le F[\omega]$. Ignoring MG for now and
taking the simplest case where $\chi$ is constant, the fundamental
eigenfrequencies of the LAWE (\ref{eq:LAWE}) satisfy
\begin{equation}\label{eq:GRstab}
\omega_0^2\le \frac{\int_0^R\dd r\,3r^2(3\Gamma_{1,0}-4)P_0}{\int_0^R\dd
r\,\rho_0r^4}.
\end{equation}
When the right hand side is negative we have $\omega_0^2<0$ and the
eigenfunctions have growing modes. This is the well-known result in stellar
astrophysics that stars where the adiabatic index falls below $4/3$ are unstable
to linear perturbations and cannot exist\footnote{Corrections from general
relativity increase this critical value to $4/3+\mathcal{O}(1) GM/R$
\cite{Chandrasekhar:1964zz}, where the $\mathcal{O}(1)$ factor
depends on the specific composition of the star. We are interested in the
properties of main-sequence stars with $GM/R\sim 10^{-6}$ and Cepheid stars with
$GM/R\sim 10^{-7}-10^{-8}$ and so this correction is always negligible compared
with the effects of MG, which are of the same order as the non-relativistic
contribution when the star is unscreened.}.

In modified gravity, we have
\begin{equation}\label{eq:modstab}
 \omega_0^2\le\frac{\int_0^R\dd
r\,3r^2\left[(3\Gamma_{1,0}-4)P_0\right]+\frac{\beta\rho_0}{\mpl}\left(r^4\frac{
\dd^2\phi_0}{\dd r^2}+2r^3\frac{\dd\phi_0}{\dd r}\right)}{\int_0^R\dd
r\,\rho_0r^4}
\end{equation}
and so this stability condition is altered in stars which are at least
partially unscreened. This is not surprising given the form of equation
(\ref{eq:MLAWE}). The term proportional to the derivative of $[(3\gao-4)P_0]$
behaves like a position dependent mass for $\xi$, which is negative when
$\gao<4/3$ so that one would expect growing modes. The additional terms in
(\ref{eq:modstab}) are due to the two new terms in (\ref{eq:MLAWE}), which are
of precisely the same varying mass form with the opposite sign. A negative mass,
which would signify an instability, coming from the GR term can then be
compensated by the new terms in MG, restoring stability\footnote{The reader may
wonder why the star is more stable in MG when the stability criterion in GR
does not change if one changes the value of $G$. When the star is unscreened we
have $G\rightarrow G(1+\alpha)$ and so one may expect any MG
effects to vanish in this limit. In GR, there is an exact cancellation coming
from the perturbations to the momentum equation (\ref{eq:momentumgen}) and the
Newtonian potential. In MG, the additional gravitational force is not derived
from the Newtonian potential but from the field profile and so any cancellation
must come from the perturbation to the field equation. A priori, there is no
reason why the field perturbations should cancel this new contribution and,
indeed, we see here that they do not.}.

Physically, the adiabatic index is a measure of how the pressure responds to a
compression of the star. Given a compression from one radius $R_1$ to a smaller
radius $R_2$, a larger adiabatic index will result in more outward pressure.
If this increase in pressure is faster than the increase in the gravitational
force, the star can resist the compression and is hence stable. Below the
critical value of $4/3$, the converse is true and the star is unstable. We have
already seen above that the new terms contributing to the stability correspond
to a varying value of $G$ and its derivative in the outer layers. Since
modified gravity enhances the gravity, one may n\"{a}ively expect that its
effect is to destabilise stars, however, we will argue below that this is not
the case. The MLAWE describes acoustic waves propagating in the star. If the
gravity and its gradient is larger in the outer layers then it is more
difficult for these waves to propagate and hence modes which would usually have
been unstable are stabilised.

Once again, we must disentangle the effects of the modified equilibrium
structure and the perturbations on the critical value of $\gao$. Consider first
the modified equilibrium structure only. In this case, the stability condition
is given by the GR expression, equation (\ref{eq:GRstab}), however the pressure
and density profiles will be different. Clearly, the critical value for the
instability is still $4/3$ since this is the only value which makes the
integral vanish but this does not necessarily mean the stability is altered
away from this value. Scaling the pressure and density using the dimensionless
quantities defined in (\ref{eq:dimlessP}), (\ref{eq:dimlessrho}) and
(\ref{eq:dimlessr}), we have
\begin{equation}\label{eq:equildestab}
 \omega_0^2\le(3\gao-4)\frac{GM}{R^3}f(\chi_0),
\end{equation}
 where $f$ is a dimensionless function which depends on the composition of
the star\footnote{The reader should note that this is a varying function of
$\chi_0$ and so is not universal.}. In MG, the radius of the star will be
smaller than its GR counterpart and the effective value of $G$ is larger.
Hence, when $\gao>4/3$ the maximum possible frequency is greater than in GR
whereas when $\gao<4/3$ the maximum frequency is more negative. If a star is
unstable in GR then MG enhances the instability, moving the frequency further
away from zero. This can also be seen from the scaling relation
(\ref{eq:backgroundomegascaling}). If $\omega_{\rm GR}^2<0$ then $\omega_{\rm
MG}^2$ is even more negative. At the background level, the effects of MG are to
destabilise stars that are already unstable, without altering the stability
condition.

Let us now turn our attention to the effects of perturbations. Using equation
(\ref{eq:mgcham}), we have
\begin{equation}\label{eq:mgstab2}
 \omega_0^2\le\frac{\int_0^R\dd
r\,3r^2\left[(3\Gamma_{1,0}-4)P_0\right]+\int_{\rs}^{R}4\pi\alpha
Gr^4\rho_0^2}{\int_0^R\dd
r\,\rho_0r^4}.
\end{equation}
The additional term is clearly positive and so one may lower the value of
$\gao$ below $4/3$ and still find positive eigenfrequencies, confirming our
earlier intuition that the effect of MG is to stabilise stars compared with GR.
Unlike GR, there is no universal critical index in MG and a numerical treatment
is necessary to extract the value for a star of given composition and screening
radius. We will investigate the stability of some simple semi-analytic models in
section \ref{sec:stab_LE} but before doing so, one can gain some insight
into the full effects of MG on the stability by using the same scaling relations
as section \ref{sec:scaling}. In particular, we can set the numerator in
(\ref{eq:mgstab2}) to zero to find the modified critical index:
\begin{equation}\label{eq:newgammacrit}
 \gao^{\rm
critical}=\frac{4}{3}-g(\chi_0)\alpha,
\end{equation}
where $g$ is a dimensionless number which encodes the effects of the
structure and composition of the star. In $f(R)$ theories, $\alpha = 1/3$
\cite{Brax:2008hh} and so we expect the critical value of $\gao$ to
change by $\mathcal{O}(10^{-1})$ assuming that $g\sim\mathcal{O}(1)$. We will
verify numerically that this is indeed the case for a simple model in
section \ref{sec:stab_LE}.

\section{Numerical Results}\label{sec:numerics}

We will now proceed to solve the MLAWE for various different stars. We will do
this for two different stellar models: Lane-Emden models, which describe the
equilibrium configuration of spheres of gas collapsing under their own gravity,
and {\scriptsize MESA} models, which are fully consistent and physical models
that include effects such as nuclear burning networks, metallicity, convection,
energy generation and time-evolution. The first models are simple compared to
the second but they have the advantage that the non-gravitational physics (e.g.
nuclear burning) is absent, which will allow us to gain a lot of physical
intuition about the new modified gravity features. They are simple
semi-analytic models and this allows us to first investigate the MLAWE
using a controlled system with known scaling properties and limits without
the complications arising from things like radiative transfer. This also allows
us to test that the code is working correctly since we can compare our results
with both the GR case, which has been calculated previously, and the fully
unscreened case, which can be predicted analytically given the GR one. Their
perturbations can also be described using an arbitrary value of $\gao$,
independent of their composition and so we will use them to study the
modifications to stellar stability. These models are not realistic enough to
compare with observational data and the power of {\scriptsize MESA} lies in that
it can produce realistic models of stars such as main-sequence and Cepheid
stars, which will allow us to predict the effects of MG on realistic stars in
unscreened galaxies. {\scriptsize MESA} predictions have already been used to
obtain the strongest constraints on chameleon-like models \cite{Jain:2012tn} and
combining these models with the modified gravity oscillation theory has the
potential to provide new constraints.

The details of the implementation of MG into {\scriptsize MESA} is given in
Appendix A (see also \cite{Davis:2011qf,Jain:2012tn} for some previous uses of
the same code) and, where necessary, details of the numerical procedure used to
solve the MLAWE are given in Appendix B. The shooting method has been used to
solve the MLAWE in all instances.

\subsection{Lane-Emden Models}\label{sec:LEmodels}

Lane-Emden models describe spheres of gas that support themselves against
gravitational collapse by producing an outward pressure given by the solution
of the (in our case modified) hydrostatic equilibrium equation. The key
assumption is that the radiation entropy per unit mass is constant, which
decouples the effects of nuclear burning (\ref{eq:engen}) and radiative transfer
(\ref{eq:radtrans}) from the stellar structure equations derived in section
(\ref{sec:eqstruc}). In this case, only the mass-conservation equation
(\ref{eq:masscons}) and the hydrostatic equilibrium equation (\ref{eq:MGHSE})
are relevant. In GR, these equations are self-similar, which allows one to scale
out the physical quantities in favour of dimensionless ones but this
property is lost if one assumes a modified gravity profile of the form
(\ref{eq:gprof}) and so, in order to retain a comparison with the GR case, we
will solve the modified hydrostatic equilibrium equation
\begin{equation}
 \frac{\dd P_0}{\dd r}=\frac{GM(r)\rho_0(r)}{r^2}\left\{
  \begin{array}{l l}
    1 & r< \rs\\
    (1+\alpha) &  r > r_{\rm s}\\
  \end{array}\right. ,
\end{equation}
which physically corresponds to setting $G\rightarrow G(1+\alpha)$ in the region
exterior to the screening radius. Lane-Emden models describe \textit{polytropic
gases} defined by
\begin{equation}
 P_0(r)= K\rho_0^{\frac{n+1}{n}},
\end{equation}
where $K$ is constant for a star of given mass but varies between different
stars. $n$, is known as the \textit{polytropic index}. The case $n=3$ was
studied in \cite{Davis:2011qf}; here we shall generalise their procedure to
arbitrary values. We can eliminate $r$ in terms of the Lane-Emden coordinate
$y=r/r_{\rm c}$\footnote{Conventionally, $\xi$ is used to represent the
Lane-Emden coordinate, however we have already used this for the radial
perturbation and so here we use $y=r/r_{\rm c}$ instead. This coordinate is
identical to the one used in \cite{Davis:2011qf} when $n=3$.} where $r_{\rm c}$
is defined by
\begin{equation}\label{eq:rc}
 r_{\rm c}^2 \equiv \frac{(n+1)P_{\rm c}}{4\pi G\rho_{\rm c}^2},
\end{equation}
where $P_{\rm c}$ and $\rho_{\rm c}$ are the central pressure and density
respectively. Next, we
re-write the pressure and density in terms of a dimensionless function $\thn(y)$
such that $P_0 = P_{\rm c}\thn^{n+1}$ and $\rho_0=\rho_{\rm c}\thn^n$.
Substituting this into the mass conservation and modified hydrostatic
equilibrium equations (equations (\ref{eq:masscons}) and (\ref{eq:MGHSE})
respectively) we can combine to two to find the modified Lane-Emden Equation:
\begin{equation}\label{eq:MGLE}
\frac{1}{y^2} \frac{ \dd}{\dd y}\left[ y^2 \frac{\dd \thn}{\dd
y}\right]=-\left\{
\begin{array}{l l}
 (1+\alpha) \thn^{n},&    y>\ys,   \\
  \thn^{n},& y<y_{\rm s},  
 \end{array}\right. ,
\end{equation}
where $\ys$ is the Lane-Emden screening radius such that $\rs=r_{\rm c}\ys$.
The boundary conditions for this equation are $\thn(0)=1$ ($P_0(0) = P_{\rm
c}$) and $\dd \thn/\dd y(0)=0$ (this follows from the fact that $M(0)=0$ in the
hydrostatic equilibrium equation). The radius of the star in Lane-Emden
coordinates, $y_R$ is then found from the condition $\thn(y_R)=0$ (the pressure
is zero at the stellar radius), from which the physical radius, $R=r_{\rm
c}y_R$ may be found. For convenience, we introduce the quantities
\begin{align}
 \omr &\equiv -y_R^2\left.\frac{\dd \thn}{\dd y}\right\vert_{y=y_R}\quad
\textrm{and} \\
\oms & \equiv -y_{\rm s}^2\left.\frac{\dd \thn}{\dd y}\right\vert_{y=y_{\rm s}}.
\end{align}
In GR, $\alpha=\chi_0=0$ and there is a unique solution for any given value of
$n$. We will denote the GR values of $y_R$ and $\omr$ using $\bar{y}_R$ and
$\bar{\omega}_R$ respectively. In MG, there is a two-dimensional space of
solutions at fixed $n$ given by specific values of $\chi_0$ and $\alpha$, each
with different values of $\omr$, $\oms$, $y_{\rm s}$ and $y_R$.

By writing equation (\ref{eq:chiint}) in Lane-Emden variables, we find an
implicit relation for $\oms$ and hence the screening radius:
\begin{equation}\label{eq:X}
 \frac{\chi_0}{GM/R}\equiv X = \left[\frac{y_R\thn(y_{\rm
s})+\omr-\frac{y_R}{y_{\rm s}}\oms}{\omr+\alpha\oms}\right],
\end{equation}
where
\begin{equation}\label{eq:LEMASS}
 M = \int_0^R4\pi r^2\rho_0(r)\dd r = 4\pi r_{\rm c}^3\rho_{\rm
c}\left[\frac{\omr+\alpha\oms}{1+\alpha}\right].
\end{equation}


As $\rs\rightarrow0$, the star becomes increasingly
unscreened, $\oms\rightarrow0$ and $\thn(y_{\rm s})\rightarrow1$. This gives the
maximum value of $X$ where the star is partially screened. For values greater
than this, equation (\ref{eq:X}) has no solutions and the star is always
unscreened. From (\ref{eq:X}), we have
\begin{equation}
 X_{\rm max} = \frac{y_R+\omr}{\omr}
\end{equation}
independent of $\alpha$. Later on, we will specify to the case $n=1.5$ so for
future reference we note here that $X_{\rm max} \approx 2.346$.

\subsubsection{Perturbations of Lane-Emden Models}\label{sec:LEperts}

We solve the MLAWE by first tabulating solutions of the modified
Lane-Emden equation and using this to numerically solve the MLAWE. This is
achieved using the shooting method and the full numerical details are given in
appendix B. The dimensionless eigenvalues 
\begin{equation}\label{eq:omtild}
 \tilde{\omega}^2\equiv \frac{(n+1)\omega^2}{4\pi G \rho_{\rm c}}
\end{equation}
for Lane-Emden models in GR were numerically calculated in 1966 by
\cite{Hurley:1966} and so as a code comparison, we have compared our fundamental
frequencies and first overtones with theirs for different values of $n$ and
$\gao$. Their values are given to five decimal places and in each case our
results matched with theirs to this accuracy.

Despite our approximation of a constant enhancement of $G$ in the region
exterior to $\rs$, self-similarity is not preserved completely in MG and so one
cannot compare solutions for fixed $\chi_0/(GM/R)$. Given a star in GR of mass
$M$ and radius $R$, one must decide upon the correct comparison in MG. In what
follows, we will fix the mass and composition (this implies that $K$ is fixed)
of the star and allow the radius to vary so that the stars we compare are stars
of the same mass whose radii (and pressure and density profiles) have adjusted
to provide an equilibrium configuration given a specific value of $\chi_0$ (and
hence $\rs$)\footnote{This is not possible in the case $n=3$ since there is
no mass-radius relation. The absence of such a relation has the result
that the constant $K$ must vary as a function of stellar mass, $\chi_0$ and
$\alpha$. For this reason, there is no meaningful way to compare perturbations
of stars in MG since one is comparing stars with different equations of state.
For this reason, we will not consider perturbations of these models.}. In order
to fix the mass, one must fix the central density and so in MG we have
$\rho_{\rm c}=\rho_{\rm c}(\chi_0,\alpha)$, highlighting the consequences of
breaking self-similarity. For concreteness, we will work with $n=1.5$. This is a
good approximation to stellar regions which are fully convective (see
\cite{kippenhahn1990stellar}, sections 7 and 13 for more details) and hence have
physical applications to red giant and Cepheid stars. In terms of an equation of
state of the form (\ref{eq:eosgam}), this model corresponds to $\gamma=5/3$. In
what follows, we will assume that $\gamma$ is identical to the adiabatic index
and set $\gao=5/3$, however, we will relax this assumption when considering
stellar stability and allow for more general models.

Using equation (\ref{eq:LEMASS}), we have
\begin{equation}\label{eq:rhocmg}
 \rho_{\rm c}=\left(\frac{M}{4\pi}\right)^2\left[\frac{8\pi
G}{2K}\right]^{3}\left(\frac{1+\alpha}{\omr+\alpha\oms}\right)^2,
\end{equation}
which may be used to find the modified Mass-Radius-$\chi_0$ relation (as
opposed to the Mass-Radius relation found in GR).
\begin{equation}\label{eq:MGR-M}
 R=\frac{1}{\left(4\pi\right)^{\frac{3}{2}}}\left(\frac{5K}{2G}
\right)\left(\frac{\omr+\alpha\oms}{1+\alpha}\right)^{\frac{1}{3}}.
\end{equation}
In the cases of red giant stars and low-mass Cepheids, we have
$GM/R\sim10^{-7}$ and so we can pick $M=M_{\rm GR}$ and $R=R_{\rm GR}$ in the GR
case ($\alpha=0$) such that $GM_{\rm GR}/R_{\rm GR}=10^{-7}$ and
\begin{equation}
 \frac{GM}{R}=10^{-7}\left(\frac{M_{\rm GR}}{M}\right)\left(\frac{R}{R_{\rm
GR}}\right)
\end{equation}
and equation (\ref{eq:X}) is (fixing $M=M_{\rm GR}$)
\begin{equation}\label{eq:scrfind}
\frac{\chi_0}{10^{-7}}=
\frac{\bar{y}_R}{y_R}\left[\frac{y_R\thn(y_{\rm
s})+\omr-\frac{y_R}{y_{\rm
s}}\oms}{\left(\bar{\omega}_R(1+\alpha)\right)^{\frac{1}{3}}
\left(\omr+\alpha\oms\right)^{ \frac{2}{3} }} \right ].
\end{equation}
The procedure is then as follows: Given a specific value of $\chi_0$, we use
a trial value of $y_{\rm s}$ to solve the Modified-Lane Emden equation until
equation (\ref{eq:scrfind}) is satisfied. We then use the Lane-Emden solution
in the MLAWE to numerically calculate the value of $\tilde{\omega}^2$ given a
value of $\gao$. Using equation (\ref{eq:rhocmg}), we can find the ratio
of the period in MG to that predicted by GR:
\begin{equation}\label{eq:taurat}
\frac{\tau_{\rm MG}}{\tau_{\rm GR}}=\sqrt{\frac{\tilde{\omega}^2_{\rm
GR}}{\tilde{\omega}^2_{\rm
MG}}}\frac{\omr+\alpha\oms}{\bar{\omega}_{R}(1+\alpha)}.
\end{equation}
We can then calculate this ratio for any values of $\chi_0$ and $\alpha$. 

Before presenting the numerical results, it is worth noting that the
fully-unscreened behaviour of the star, at least in the case where only the
effects of the modified equilibrium structure are considered, can be calculated
in terms of the GR properties of the star. In the fully-unscreened case, one
has $\oms=y_{\rm s}=0$. One can then set $y\rightarrow
(1+\alpha)^{-\frac{1}{2}}y$ to bring equation (\ref{eq:MGLE}) into the same form
as in GR. This then gives $y^{\rm unscreened}_R =
(1+\alpha)^{-\frac{1}{2}}\bar{y}$ and $\omr^{\rm
unscreened}=(1+\alpha)^{-\frac{1}{2}}\bar{\omega}_R$. For $n=1.5$, one has
$\bar{y}_R\approx 3.654$ and $\bar{\omega}_{\rm R}=2.72$. One then has, by
rescaling the MLAWE (see Appendix B for the equation in these coordinates),
$\tilde{\omega}^2_{\rm{ unscreened}}/\tilde{\omega}_{\rm GR}^2=(1+\alpha)$ and
so, using (\ref{eq:rhocmg}), $\omega^2_{\rm{ unscreened}}/\omega_{\rm
GR}^2=(1+\alpha)^4$, which exactly matches our prediction in
(\ref{eq:backgroundomegascaling}) for $\gamma=5/3$. From equation
(\ref{eq:taurat}), we have $\tau^{\rm unscreened}_{\rm MG}/\tau_{\rm
GR}=(1+\alpha)^{-2}=0.5625$ for $\alpha=1/3$. These unscreened results can be
used to check that the numerical results are behaving as expected. 

We would like to investigate the effect of the different modifications coming
from the altered equilibrium structure and the modified perturbation equation
separately. They appear at the same order in the MLAWE and so we expect them to
contribute equally and it is important to dis-entangle their effects,
especially since we have already argued in section \ref{sec:stellarstability}
that they may contribute differently to the stellar stability and that the
equilibrium structure acts to make negative GR frequencies more negative.
Lane-Emden models are perfectly suited for this study since we do not have to
worry about altered evolution histories and so we will consider two cases:
\begin{itemize}
\item case 1: We solve the LAWE (\ref{eq:LAWE}) using modified Lane-Emden
profiles. This case only includes the modified equilibrium structure.
\item case 2: We solve the full MLAWE using both the modified background
structure and the modified perturbation equations. This is the physically
realistic case.
\end{itemize}
The case where we ignore the modified equilibrium structure and include only the
perturbations is highly unphysical, there is no screening radius and so we are
introducing the perturbations about an arbitrary radius. Furthermore, the size
of the effect depends on whether we take the radius of the star as
corresponding to the GR solution (in which case we are ignoring the effects on
the period coming from the change in the critical density (\ref{eq:rhocmg})) or
the equivalent MG solution (in which case our profiles do not satisfy the
boundary conditions and we are including some of the modified equilibrium
properties in our analysis). For these reasons, we do not investigate this
scenario.  In each case, we assume the profile (\ref{eq:gprof}). Since we have
solved the modified Lane-Emden equation with constant $\alpha$ in the region
$r>\rs$ (this is required for a physically meaningful comparison with GR) this
profile is not technically correct since it does not satisfy the stellar
structure equations. In fact, it is an under-estimate \footnote{Compared to the
result that would be obtained if we had used a fully-unscreened profile. The
equilibrium structure is still a small over-estimate of the effects of MG. For
our purposes, this is not an issue since we seek only to investigate the new
effects of MG oscillations and do not compare any of these results with real
stars. When we analyse {\scriptsize MESA} predictions we will use a fully
consistent approach.}. In each case we will fix $\alpha=1/3$, corresponding to
$f(R)$ gravity and vary $\chi_0$.

\begin{figure}
\includegraphics[width=0.5\textwidth]{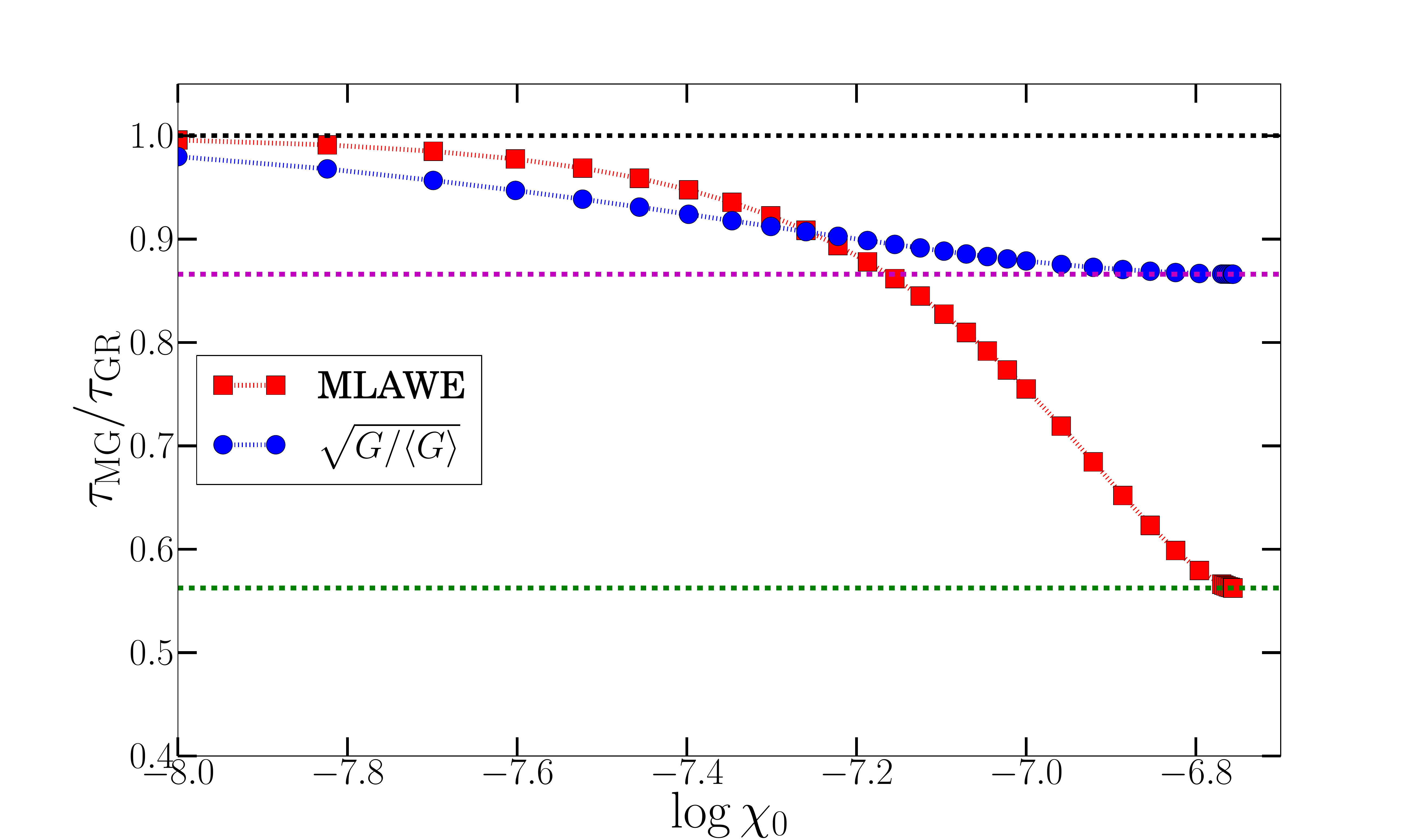}
\caption{The fractional change in the stellar pulsation period as a function of
$\log_{\rm 10}\chi_0$ when only the change to the equilibrium structure is
considered (case
1). The red squares correspond to eigenfrequencies of the LAWE whereas the blue
circles show the approximation (\ref{eq:periodapprox}). The green dashed line
shows the ratio for a fully unscreened star and the black dashed line shows a
ratio of 1, corresponding to a GR star. The magenta line shows the
fully-unscreened value of $\sqrt{\frac{G}{\langle
G\rangle}}=(1+\alpha)^{-\frac{1}{2}}$.}\label{fig:noperts}
\end{figure}

In figure \ref{fig:noperts} we plot the ratio of the MG period to the GR
one as a function of $\log\chi_0$ for case 1. In a previous study
\cite{Jain:2012tn}, the approximation \begin{equation}\label{eq:periodapprox}
\frac{\tau_{\rm MG}}{\tau_{\rm GR}}=\sqrt{\frac{G}{\langle G\rangle}},
\end{equation}
where $\langle G\rangle$ is the average value of the effective Newtonian
constant using some appropriate weighting function was used in order to obtain
new constraints on the model parameters. This approximation is based on the
change in the equilibrium structure only and so it is important to test not only
how it compares with the predictions from the full numerical prediction at the
background level but also how well it can be used to approximate the frequency
once MG perturbations are taken into account. Hence, we also plot the
approximation using the Epstein \cite{1950ApJ...112....6E} weighting function as
was used by \cite{Jain:2012tn}. In each case we have calculated $\langle
G\rangle$ using the modified Lane-Emden solution at given $\chi_0$. One should
emphasise that \cite{Jain:2012tn} used this approximation for {\scriptsize
MESA} models whereas this comparison is purely for the hypothetical case of
Lane-Emden models. We will investigate how well this holds for {\scriptsize
MESA} models in section \ref{sec:MESAperts}. The figure reveals that the
approximation (\ref{eq:periodapprox}) is an over-estimate for very screened
stars whereas it is a large under-estimate for stars that are significantly
unscreened. The Epstein function favours the regions of Cepheid stars that are
most important for pulsations. This tends to be the outer layers and so it is
no surprise that it over-estimates the effects in the screened case: it places
a large emphasis on the small region where the gravity is enhanced even though
this region has little to no effect on the structure of the star. The
approximation (\ref{eq:periodapprox}) assumes that the stellar radius is fixed
but in Lane-Emden models this is clearly a decreasing function of
$\chi_0$ and $\alpha$ according to (\ref{eq:MGR-M}). According
to (\ref{eq:dimOmega}), the period scales as $R^{\frac{3}{2}}$, which explains
why the approximation is an underestimate when the star is very unscreened and
the change in the radius is significant. We can also see that the ratio
asymptotes to the value predicted above when the star is fully-unscreened.

\begin{figure}
\includegraphics[width=0.5\textwidth]{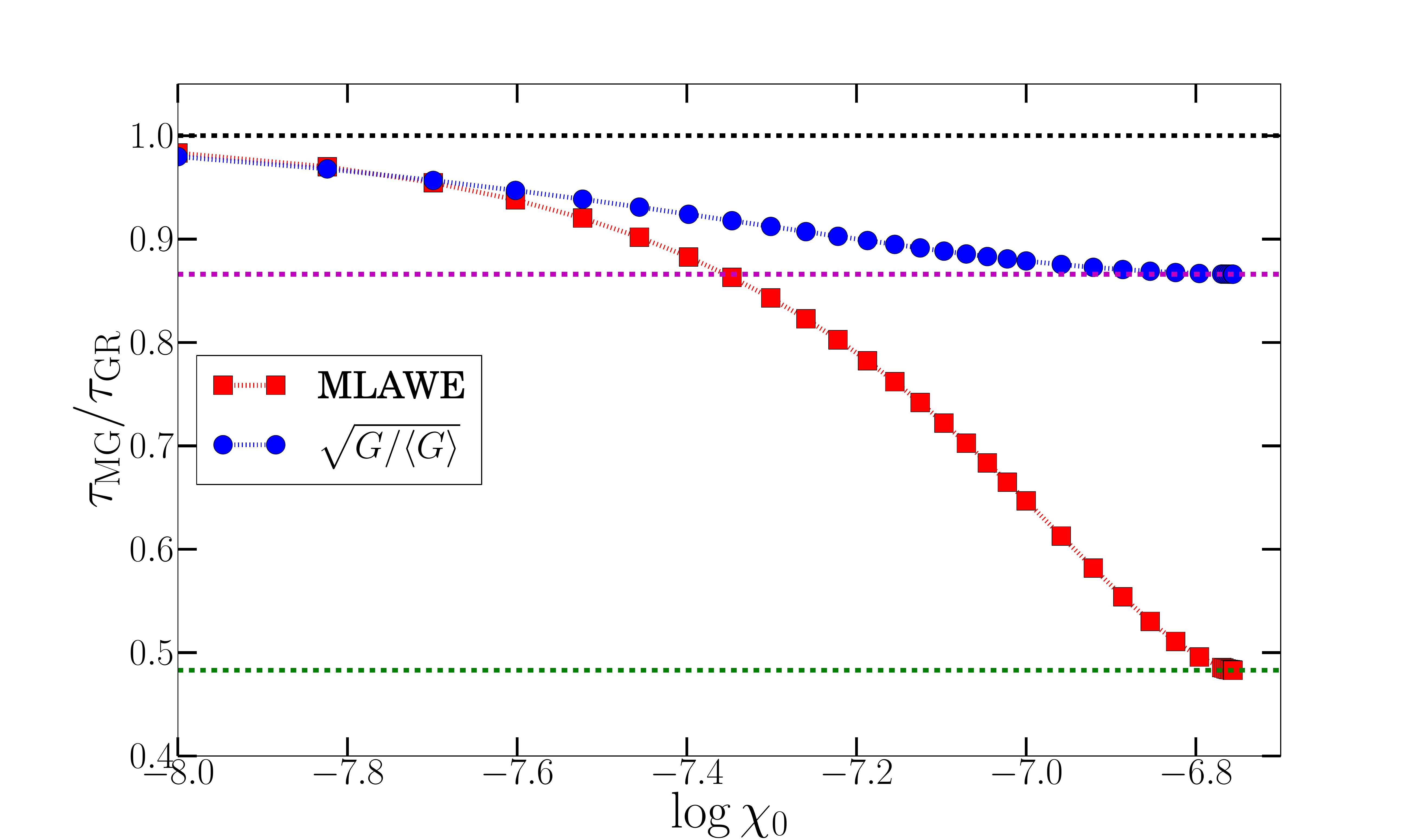}
\caption{The fractional change in the stellar pulsation period as a function of
$\log_{\rm 10}\chi_0$ when both the effects of the modified equilibrium
structure and
modified perturbation equation are considered (case 2). The red squares
correspond to eigenfrequencies of the MLAWE whereas the blue circles show the
approximation (\ref{eq:periodapprox}). The green dashed line shows the ratio for
a fully unscreened star and the black dashed line shows a ratio of 1,
corresponding to a GR star. The magenta line shows the fully-unscreened value
of $\sqrt{\frac{G}{\langle
G\rangle}}=(1+\alpha)^{-\frac{1}{2}}$.}\label{fig:case3}
\end{figure}

In figure \ref{fig:case3} we plot the ratio of the MG period to the GR one for
case 2. We can see that the approximation (\ref{eq:periodapprox}) fails very
rapidly and that the change in the period is significant and can be as large
as 50\% for significantly unscreened stars.

\begin{figure}
\includegraphics[width=0.5\textwidth]{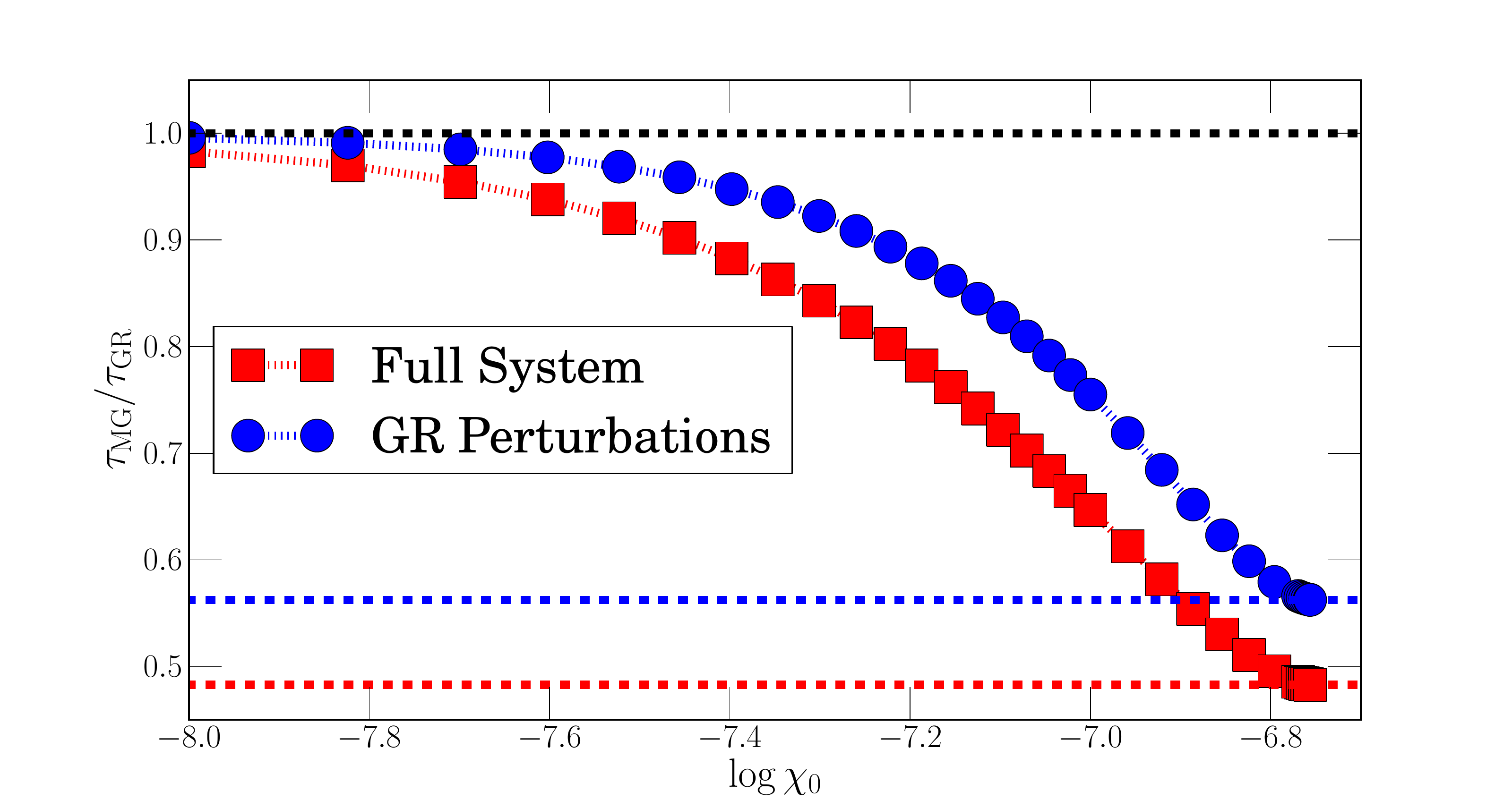}
\caption{The fractional change in the stellar pulsation period as a function of
$\log_{\rm 10}\chi_0$ for case 1 (blue circles) and case 2 (red
squares). The black dashed line shows the GR ratio of $1$ and the red and blue
lines show the fully-unscreened ratio for the
full simulation and the one including only the modified
equilibrium structure respectively.}\label{fig:all3}
\end{figure}

We then plot the two cases together in figure \ref{fig:all3}. One can see that
the effect of the new terms coming from the modified structure of hydrodynamics
has a significant effect on the period and that if one were to consider only
the background structure, the change in the period would be a large
under-estimate. That being said, the change in the period from the GR value is
$\mathcal{O}(1)$ as soon as one calculates using the modified equilibrium
structure and the effect of the perturbation is to increase this by an amount
not as large as this initial change. These results seem to suggest that
convective stars such as Cepheid and red giants may show very large changes in
the oscillation periods due to their modified background structure and that the
approximation will tend to under-estimate this change. Furthermore, the effects
of the hydrodynamic perturbations will make these changes more drastic but not
as large as those coming from the modified background. In fact, we will see
below that this is not the case for Cepheid models. We will see that the
approximation holds very well when only the background structure is considered
but when the perturbations are included the resulting change in the period is
three times as large as that due to the modified equilibrium structure alone.
One assumption we have made here is that the constant $K$ appearing in the
polytropic relation is constant. This is tantamount to having a uniform
composition throughout the entire star. This is a good approximation for red
giant stars, which are fairly homogeneous but Cepheids have shells of varying
composition and several ionised layers and so this is likely not an accurate
approximation. In particular, we will see below that the radius of Cepheid stars
does not change significantly in MG, contrary to what this model would predict
and this is why we find the approximation holds well despite this models
predictions. 

\subsubsection{Stability of Lane-Emden Models}\label{sec:stab_LE}

Before moving on to look at realistic models from {\scriptsize MESA}, we will
first use Lane-Emden models to investigate the modification to the stellar
stability criterion. In section \ref{sec:stellarstability} we derived the new
properties relating to stellar stability in MG and argued two new features:
first, that when there are unstable modes present in GR such that
$\omega_0^2<0$, then, when only the change equilibrium structure is taken into
account, the instability is worse i.e. $\omega_0^2$ is more negative;
second, that the new term appearing in the MLAWE makes stars more stable, 
the critical value of $\gao$ required for $\omega_0^2<0$ is less
than the GR value of $4/3$ and the correction is of order
$g(\chi_0)\alpha$ given in (\ref{eq:newgammacrit}). $g(\chi_0)$ encodes the
competing effects of the new term in the MLAWE and the modified structure and
composition coming from the new equilibrium structure. Here, we will verify
these predictions numerically.

\begin{figure}
\includegraphics[width=0.5\textwidth]{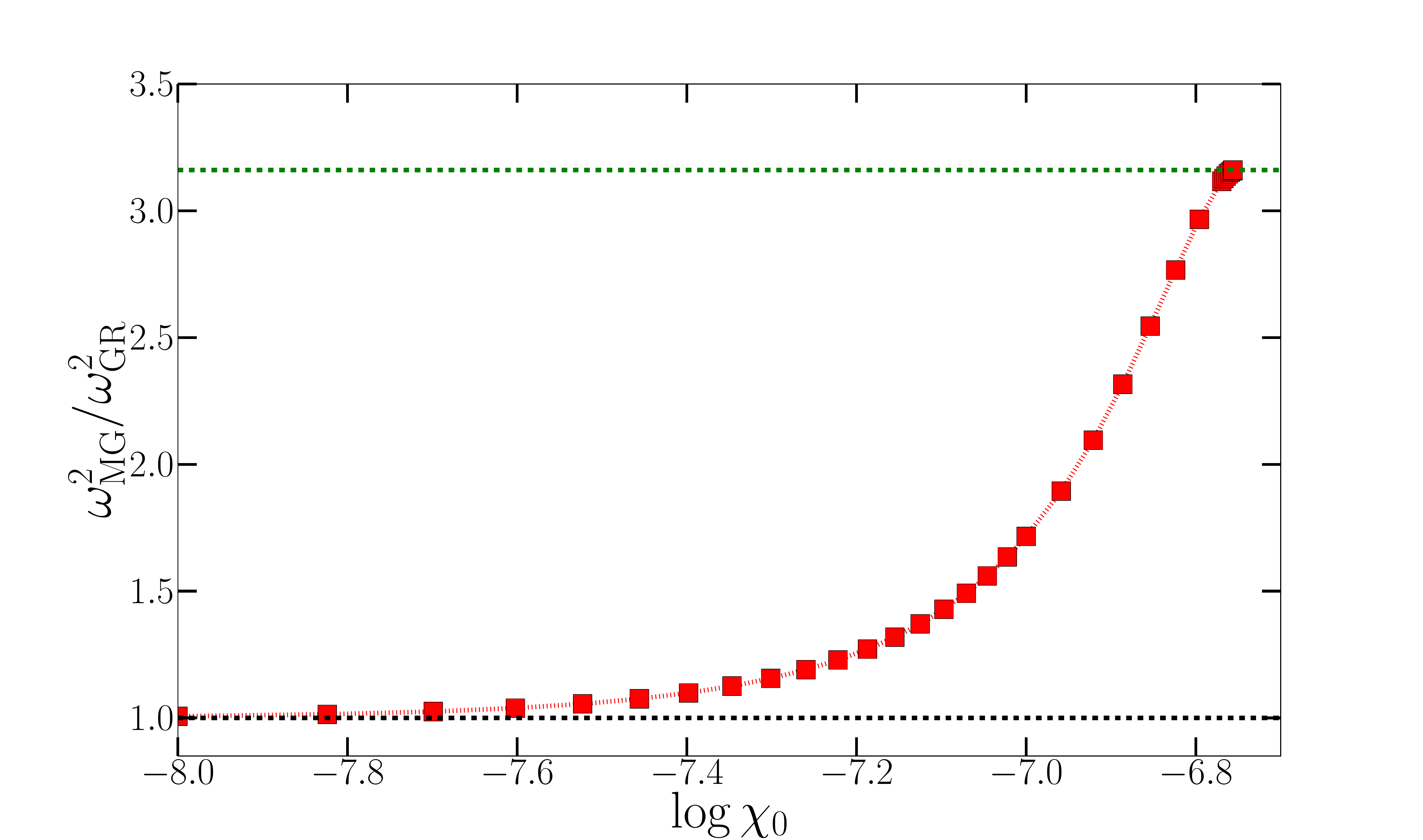}
\caption{The ratio of the MG to GR eigenfrequencies as a function of $\log_{\rm
10}\chi_0$ when $\gao\approx 1.2333$.}\label{fig:gl0}
\end{figure}

In order to investigate the first, we have solved for the modified
eigenfrequencies of the same $n=1.5$ modified Lane-Emden model investigated in
the last subsection ignoring the new term in the MLAWE and using various values
of $\gao<4/3$. This corresponds to a star whose adiabatic perturbations are
governed by a different index to that appearing in the equation of state that
fixes the equilibrium structure. In each case, the modified eigenfrequencies are
indeed more negative the more unscreened the stars are and, as an example, we
plot the ratio $\omega^2_{\rm MG}/\omega^2_{GR}$ in the case
$\gao=37/30\approx1.23333$, which is close to being stable. In this case one has
$\omega^2_{\rm GR}=-0.314$ and so the larger this ratio, the more negative the
MG value. This is plotted in figure \ref{fig:gl0} and it is evident that the
instability is indeed worse in stars which are more unscreened.

\begin{figure}
\includegraphics[width=0.5\textwidth]{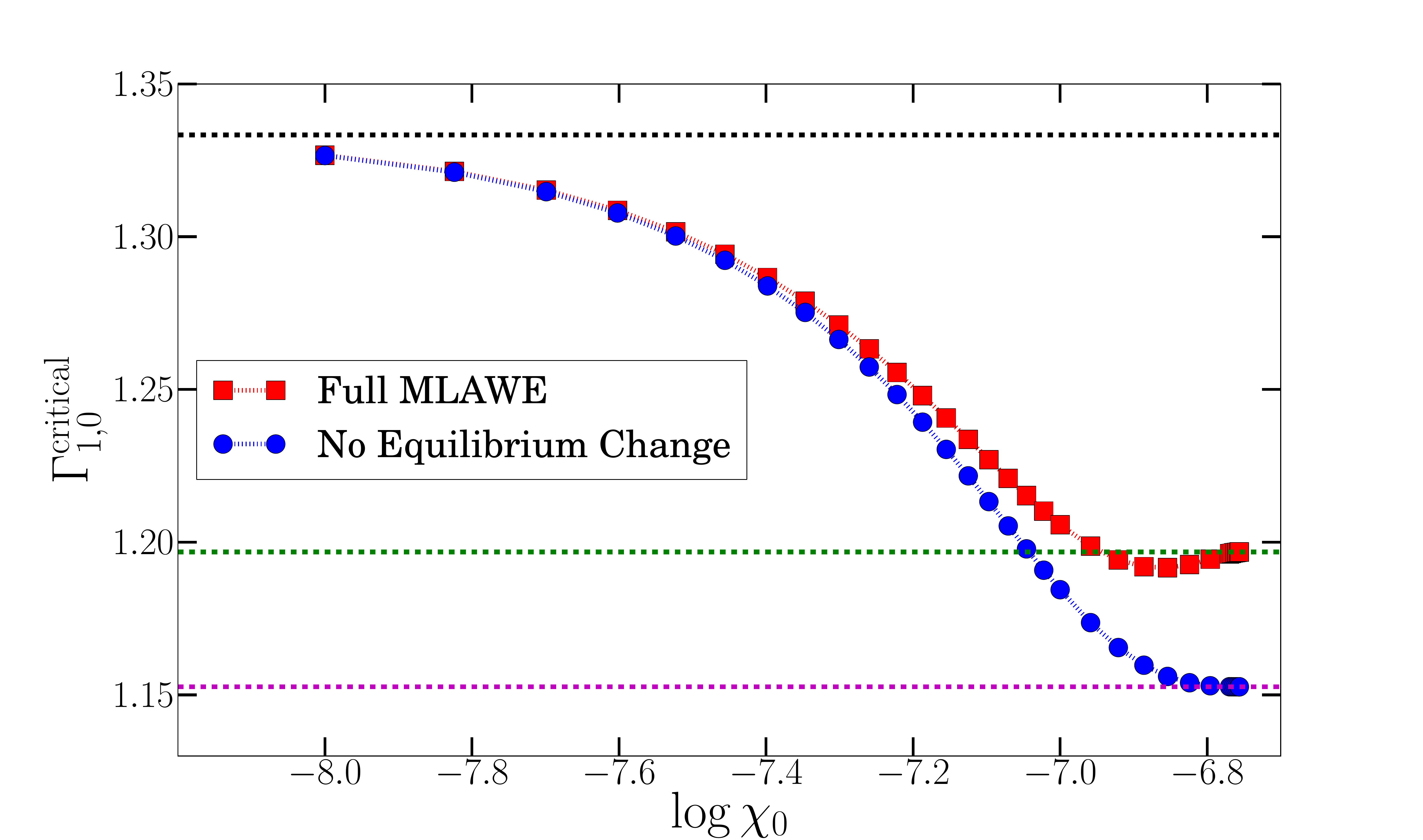}
\caption{The critical value of $\gao$ as a function of $\log_{\rm 10}\chi_0$.
The blue circles show the critical values when the modifications to the
equilibrium structure are ignored and the red squares show the critical values
when the full MLAWE is solved. The black dashed line shows the GR value of $4/3$
and the magenta and green lines show the fully unscreened values in the no
modified equilibrium and full cases respectively.}\label{fig:critgam}
\end{figure}

Next, we turn our attention to the modification of the critical value of
$\gao$. In order to investigate this, we again use the $n=1.5$ model above and
vary $\gao$ as a function of $\chi_0$. We scan through different values of
$\gao$ at fixed $\chi_0$ in order to find the value where $\omega^2\approx0$
(to 8 decimal places), which is the new critical value in MG. We solve for the
zero-eigenfrequencies in two cases: the case where we ignore the modified
equilibrium structure\footnote{We have already argued in the previous subsection
that the first case is highly unphysical and ambiguous. This is true if we
wish to discern how the MG perturbations affect the numerical value of the
oscillation periods but here we seek only to qualitatively investigate how the
modified equilibrium structure influences the critical adiabatic index. Hence,
for this purpose it is a reasonable case to investigate.} and include only the
MG perturbations and the full MLAWE. The values of the critical value of $\gao$
vs $\log\chi_0$ are plotted in
figure \ref{fig:critgam}.

It is evident from the figure that the critical value is indeed lower than
$4/3$ showing that these models are indeed more stable in MG. One can also see
that the red curve lies above the blue one so that the effect of the modified
equilibrium structure is to destabilise the star compared with how stable it
would have been had only the modified perturbation structure been present. We
have already argued this in equation (\ref{eq:equildestab}), where we found
that the contribution to the stability condition from the equilibrium structure
is larger in MG gravity. In terms of the dimensionless function $g(\chi_0)$,
(\ref{eq:equildestab}) appears as a denominator and so the effect of the
background is to reduce the correction to the GR critical index. The critical
value when the modified background structure is ignored decreases monotonically,
however, the full MLAWE predicts an increase in stars that are significantly
unscreened, showing that the effects of this altered equilibrium structure
become more important when the star is more unscreened. This is consistent with
our findings in the previous subsection. Finally, we note that the change in the
critical index is indeed of order $10^{-1}$, which we predicted using analytic
arguments in section \ref{sec:stellarstability}.

\subsection{MESA Models}\label{sec:MESAperts}

Having studied some simple stellar models and gained some intuition about the
MLAWE, we now turn our attention to realistic stellar models from {\scriptsize
MESA}, which provides consistent stellar models including the full effects of
modified gravity. The implementation of modified gravity into {\scriptsize
MESA} is discussed in appendix A. We will limit the discussion to Cepheid
stars, which are useful observational tools to study modified gravity and have
already been used to place the strongest constraints on $\chi_0$ and $\alpha$ to
date \cite{Jain:2012tn}. They are the only stars whose oscillations can be
observed in distant galaxies\footnote{Some other pulsating objects such as RR
Lyrae stars can be resolved in the local group but this is necessarily
screened and so these objects cannot be used to probe MG. One requires an
unscreened distance indicator at suitably large distances such that the dwarf
galaxy is unscreened and there is another screened distance indicator which can
be used in a comparison. }.

\subsubsection{Perturbations of Cepheid Models}

\cite{Jain:2012tn} have used the perturbed period-luminosity relation
(\ref{eq:plmg}) with the approximation $\Delta\tau/\tau\approx \Delta G/2G$,
where $\Delta G\equiv \langle G\rangle-G$ is found using the Epstein
\cite{1950ApJ...112....6E} function, to find the difference between the inferred
distances in MG and GR. We are now in a position to calculate $\Delta\tau/\tau$
for the same models which they have used and check how well this approximation
holds. In table 1 they present six models with different $\chi_0$ and $\alpha$
for a 6$M_\odot$ star with initial metallicity $z=0.004$. Using {\scriptsize
MESA}, we have recreated these models using the same procedure that they
outline. Using the instability strip given by \cite{Alibert:1999an},
\begin{align}
 \log L = 4.2 - 46 \left(\log T_e - 3.8\right)
\end{align}
for the blue edge we select first-crossing models \footnote{There is some
ambiguity in
the literature as to the precise definition of first-crossing. Here, we mean
that the star has ascended the red-giant branch and subsequently looped,
crossing the strip for the first time and not the brief, unobservable crossing
after the main-sequence.} using their different parameter values and an
additional model where the star has evolved under GR. The red-edge is determined
by a dissipation term in the LAWE due to convection, which we have not included,
and so we do not make any statements about red-edge properties. We then check
that our models agree with theirs by calculating $\Delta G/G$ and we indeed find
the same values. Next, we calculate the modified periods by solving the MLAWE
and we again examine both the full solution and the solution where we ignore the
MG perturbations in order to discern their new effects. The models with
$\alpha=1/3,\,\chi_0=10^{-6}$, $\alpha=1/2,\,\chi_0=10^{-6}$ and
$\alpha=1,\,\chi_0=4\times10^{-7}$ all execute one loop that does not cross the
instability strip and cross only on the second loop. There is hence no model to
compare to the GR first-crossing and so we do not analyse these
models\footnote{This is not to say that the approximation used by
\cite{Jain:2012tn} is not applicable, rather that numerically calculating the
period will give spurious results due to the lack of a suitable GR model to
compute the unmodified period.}.

In table \ref{tab:cepheids} we show $\Delta \tau/\tau$ calculated for each of
their models using the approximation, the case of MG perturbations turned
off and the full MLAWE. In each case, the change in the radius is $\Delta
R/R\sim\mathcal{O}(10^{-2})$ and so most of the change in the period is due to
modified gravity and not the size of the star. This is very different from the
$n=1.5$ Lane-Emden model in subsection \ref{sec:LEperts} which predicts a large
reduction in the radius when the star is unscreened. It is hardly surprising
then that the approximation works very well when the MG perturbations are
ignored; this approximation was found by perturbing the relation $\tau\propto
G^{-\frac{1}{2}}$ at fixed radius. When the MG perturbations are included, we
see that this approximation breaks down and the relative difference in the
period is $\mathcal{O}(10^{-1})$, which is approximately a day. Table
\ref{tab:cepheids2} shows how these changes propagate to $\Delta d/d$, which is
the astrophysical quantity used to place constraints. Again, we see that the
approximation is good when one ignores the MG perturbations but when
these are included the relative distance difference can be up to three times as
large. We therefore conclude that the constraints of \cite{Jain:2012tn} are
conservative, and it is possible that they could be improved given the larger
change predicted in the full theory.

With this in mind, we estimate the values of $\chi_0$ that can be probed using
the full MLAWE rather than the approximation for the same values of $\alpha$
discussed above. We accomplish this by taking the same initial stellar
conditions and running a series of new simulations using {\scriptsize MESA} for
successively decreasing values of $\chi_0$. Using the same procedure to
identify the Cepheid models at the blue edge as described above, we calculate
$\Delta d/d$ using the MLAWE until it is equal to the value predicted by the
approximation. These values were those used in \cite{Jain:2012tn} to obtain
their constraints. In table \ref{tab:constraints} we show the values of $\chi_0$
and $\alpha$ such that the MLAWE gives the same result as the approximation.
These represent an estimate on the range of parameters that one could hope to
constrain using the same data-sets and the MLAWE. Of course, this is just a
simple estimate and a more rigorous method would be to redo their
analysis, which is beyond the scope of this work. Nevertheless, this simple
estimate goes to show that we can expect new constraints significantly stronger
than the previous ones, in particular, the MLAWE predictions suggest that the
constraints could be pushed into the $\mathcal{O}(10^{-8})$ regime.

\begin{table*}
\centering
\begin{tabular}{|c|c|c|c|c|}
\hline
$\alpha$ & $\chi_0$ & $\Delta \tau/\tau$ (approximation) & $\Delta \tau/\tau$
(no perturbations) & $\Delta \tau/\tau$ (full MLAWE)   \\
\hline
$1/3$ & $4\times10^{-7}$ & $0.086$ & $ 0.092$ & $0.266$ \\
\hline
$1/2$ & $4\times10^{-7}$ & $0.054$ & $ 0.064$ & $0.207$ \\
\hline
$1$ & $2\times10^{-7}$ & $0.102$ & $ 0.122$ & $0.314$ \\
\hline
\end{tabular}
\caption{The change in the period of Cepheid pulsations due to MG effects.}
\label{tab:cepheids}
\end{table*}
 \begin{table*}
\centering
\begin{tabular}{|c|c|c|c|c|}
\hline
$\alpha$ & $\chi_0$ & $\Delta d/d$
(approximation) & $\Delta d/d$ (no perturbations)& $\Delta d/d$
(full MLAWE)   \\
\hline
$1/3$ & $4\times10^{-7}$ & $-0.03$ & $ -0.04$ & $-0.12$  \\
\hline
$1/2$ & $4\times10^{-7}$ & $-0.05$ & $ -0.06$ & $-0.16$  \\
\hline
$1$ & $2\times10^{-7}$ & $-0.06$ & $ -0.07$ & $-0.19$  \\
\hline
\end{tabular}
\caption{The change in the inferred Cepheid distance due to MG. In each case
$\Delta d/d$ was found found using the perturbed P-L relation \ref{eq:plmg}.}
\label{tab:cepheids2}
\end{table*}
\begin{table}
\centering
\begin{tabular}{|c|c|}
\hline
$\alpha$ & $\chi_0$  \\
\hline
$1/3$ & $9\times10^{-8}$  \\
\hline
$1/2$ & $7\times10^{-8}$   \\
\hline
$1$ & $3\times10^{-8}$  \\
\hline
\end{tabular}
\caption{The lower bounds on $\chi_0$ and $\alpha$ that could potentially be
placed if one were to use the same procedure and data-sets as
\cite{Jain:2012tn} using the full MLAWE instead of the approximation.}
\label{tab:constraints}
\end{table}

\section{Discussion and Conclusion}\label{sec:concs}

We have presented a study of the properties of stellar oscillations in modified
theories of gravity such as chameleons and symmetrons. Starting with a generic
fifth-force appropriate for all couplings of a scalar to matter of the
form $\mathcal{L}_{\rm coupling}=C(\phi)T_m$, we have derived the equations of
modified gravity hydrodynamics. At zeroth-order, these give us the modified
equations of stellar structure, which have been well-studied before
\cite{Chang:2010xh,Davis:2011qf,Jain:2012tn}. At first-order, we obtain the
equations of motion describing radial and adiabatic perturbations of the star
coupled to perturbations of the scalar field. This describes scalar radiation
which is known to be small and so we ignore the coupling terms to find the
equations governing the oscillations of stars in modified gravity, the modified
linear adiabatic wave equation. This is a Sturm-Liouville problem where the
eigenvalues correspond to the angular frequency of oscillations. We have
discerned two new features due to the modification of this equation: first, the
period of oscillations is greatly reduced compared to what one would
n\"{a}ively expect from considering the modified equilibrium structure alone
and second, the adiabatic index, $\gao$, can be lowered below the GR limit of
$4/3$ while keeping $\omega_0^2$, the fundamental frequency, positive so that
there are no unstable modes. This means that stars are more stable in MG than in
GR and, unlike GR, there is no universal bound on $\gao$ so that the value where
$\omega_0^2=0$ depends on the composition of the star as well as the MG
parameters. We found that the modified equilibrium structure by itself could not
alter the stability bound but it does have an effect on the new critical value
of $\gao$ once the perturbations are included. Specifically, it raises it
closer to the GR value relative to what it would have been if we had
perturbed around the GR solution. Whilst the modified background structure
cannot change the instability criterion alone, if an instability is present then
it acts to make $\omega^2$ more negative than the GR prediction.

After discussing these features in length, we have investigated them
numerically. Using a generalisation of the procedure in \cite{Davis:2011qf}, we
have solved the MLAWE for n=1.5 Lane-Emden models, which describe convective
stars. We numerically calculated the change in the period as a function of
$\chi_0$ for $f(R)$ gravity ($\alpha=1/3$) and found that it monotonically
decreased. We found that the change in the equilibrium structure produced very
large changes from the GR values and that these increase by a significant but
smaller amount once the effects of modified gravity hydrodynamics are included.
We discussed the application of this model to real stars and concluded that
whilst likely appropriate for red giants, it is probably not a good
approximation to Cepheid stars, especially in light of the results from
{\scriptsize MESA} discussed below. Nevertheless, this is a useful model whose
simplicity allowed us to gain a lot of intuition about the behaviour of the
MLAWE and how it responds to various MG features. In particular, it revealed
how the effects of the modified hydrodynamic perturbations can be important if
one wishes to calculate the correct oscillation period. Next, we investigated
the stability properties of the models. We confirmed numerically that $\omega^2$
is indeed more negative than predicted in GR when considering only the modified
equilibrium structure. We then re-introduced the MG terms into the MLAWE and
used the same Lane-Emden models to numerically find the new critical value of
$\gao$ as a function of $\chi_0$. We indeed found that this was always less than
$4/3$ so that these stars are indeed more stable. When the equilibrium structure
is ignored so that we perturb around the GR solution,
this decrease is monotonic. When we then consider the full MLAWE, the modified
equilibrium structure acts to raise the critical value above what it would be
if we had perturbed around GR but never to $4/3$. When the star is unscreened,
the effects of the modified equilibrium structure begins to dominate and
the critical value for very unscreened stars is not as small as some which are
partially-screened.

Having gained a lot of intuition from these simple models, we turned our
attention to models coming from {\scriptsize MESA}. {\scriptsize MESA} is a
stellar structure code which can consistently simulate the structure and
evolution of stars over their entire life and includes every important stellar
process such as convection, mass-loss, nuclear burning, opacity tables etc. In
\cite{Davis:2011qf}, we presented a modified version which includes the effect
of chameleon-like theories in a model independent way (see also
\cite{Chang:2010xh} for a different implementation). This can be used to make
quantitative predictions which can be compared to data. Indeed, this has
already been done by \cite{Jain:2012tn} who used it to find approximations to
the change in the period of Cepheid pulsations, which was then used to
constrain $\chi_0$ and $\alpha$ by comparing the distances to the same
unscreened dwarf galaxies measured using Cepheid and tip of the red giant
branch techniques.

Here, we have explored how the MG perturbations affect the periods of Cepheid
pulsations. These stars are good probes of MG and here we have found that
the perturbations do indeed alter the modified periods significantly. If we
ignore the MG perturbations then our numerical calculation of $\Delta d/d$,
the change in the inferred Cepheid distance, agrees very well with the
approximation used in \cite{Jain:2012tn}. Once the perturbations were
included, we found that this change could be up to three times as large as the
approximation predicts. 

Previous studies \cite{Jain:2012tn} have used this approximation and it is the
the astrophysical errors that limit the strength of the constraint. What we have
shown here is that these constraints are conservative and it may be
possible to reduce the bounds further using the same data-sets if one were to
use the results from the full MLAWE. In particular, we have estimated that one
could probe the self-screening parameter to values of $\mathcal{O}(10^{-8})$
when $\alpha\sim\mathcal{O}(1)$, although this is by no means a rigorous
statistical analysis and a more detailed treatment is left for future work.

The altered stability condition, whilst interesting, is unlikely to be a good
probe of modified gravity in the near-future. The increased stability has very
few observational consequences. For example, one might look for physical
processes where $\gao$ falls below $4/3$. This is the case with type II
supernovae, where the progenitor stars are very massive with large radiation
pressures. Upon compression, the extra photon energy is used to
photo-disintegrate iron nuclei and very little is transferred to the pressure
so that $\gao<4/3$. One might imagine then that the type II supernova rate in
dwarf galaxies would be less if they were unscreened so that galactic processes
like supernova feedback etc. are less efficient. Unfortunately, the current lack
of complete models for type II supernovae and the fact that the feedback
efficiency (and supernova rate) tend to be packaged into efficiency parameters
(see, for example, \cite{Hatton:2003du,Sakstein:2010mh} and references therein)
that are not necessarily physical makes a quantitative prediction that can be
compared to observations difficult. That being said, one could potentially look
for stars in dwarf galaxies which are not seen in the HR diagrams of
stellar populations in our own galaxy. Such stars would likely be very massive
and hence have high luminosities so that it may be possible to
resolve them. This remains to be seen and the study of the {\scriptsize MESA}
eigenfrequencies of such models could provide some clues as to their
properties. This is left for a future investigation.

Finally, it is worth commenting on the limitations of this approach and
future work. Non-radial modes are generally not observable in distant
galaxies and so the specialisation to radial modes is sensible. We have
only studied adiabatic oscillations, however, it is well-known that Cepheid
pulsations are driven by non-adiabatic effects. Indeed, the blue edge of the
instability strip corresponds the the location in the HR diagram where the
motion becomes non-adiabatic inside the ionisation layer. This appears as a
driving term in the LAWE and the generalisation to MG is straightforward. This
driving mechanism depends only on atomic physics and thermodynamics, which are
insensitive to MG and so the MG equivalent is simply the MLAWE with exactly the
same driving term. What is different is the internal structure of the star and
it is entirely possible that the location of the blue edge is different in MG.
Furthermore, with this driving term the MLAWE is inhomogeneous and one can
derive the amplitude as well as the frequency, and so one could, in
theory, derive the modified period-mass-luminosity (P-M-L) relation in MG
without needing to perturb the GR calibrated one. This is left for future work,
however, we remark that the driving term tends to excite the fundamental
frequency, which is precisely what we have calculated here. 

We have ignored the effects of scalar radiation, which is known to be
negligible in chameleon-like theories, which have been the focus of this work.
It is still unclear whether this is the case for Galileon-like
theories and this too is left for a future investigation. It would also be
interesting to see if effects that arise in coupled systems, such as beating
patterns in the P-L relation, could be present, although this is beyond the
scope of this work.

Here, we shall be content to present the theory and underlying framework for
describing perturbations in modified gravity and to have shown that the full
analytic treatment can lead to effects far stronger than one would n\"{a}ively
expect given the modified equilibrium structure alone.

\begin{acknowledgments}I am incredibly grateful to Eugene Lim for several
helpful discussions and suggestions. This work has greatly benefited from
discussions with Avery Broderick, Phillip Chang and Bhuvnesh Jain. I am also
thankful for conversations with Mustafa Amin, Anne-Christine Davis, Tom Hogan,
Lam Hui, Will Keen, Levon Pogosian, Fabian Schmidt, Alessandra Silvestri and
Vinu Vikram. I would like to thank the anonymous referee, whose questions and
suggestions have greatly improved the quality of this paper. I am supported by
the STFC.\end{acknowledgments}

\appendix

\section{Implementation of Modified Gravity into MESA}

{\scriptsize MESA} \cite{Paxton:2010ji,Paxton:2013pj} is a publicly available
program capable of solving the complete system of stellar structure equations
coupled to the radiative transfer system, stellar atmosphere models, nuclear
burning networks, convective motion and micro-physical processes such as
opacity and electron degeneracy. It also includes effects such as
overshooting, mass-loss and rotation in a fully consistent manner. Given some
initial mass, it generates a pre-main-sequence stellar model and dynamically
evolve it through the main-sequence and subsequent post-main-sequence to its
final state, be it a white dwarf, neutron star or core-collapse supernova. Here,
we are mainly interested in the main-sequence and post-main-sequence phases and
so we shall limit our discussion to these.

{\scriptsize MESA} is a one-dimensional code (in that it assumes spherical
symmetry) that divides the star into a series of variable-length cells, each
with a specific set of quantities such as temperature, density, mass fraction
etc. that may correspond to the values at either the cell centre or
boundary. Exactly which depends on the quantity in question, however it is
always possible to interpolate between the two. We implement the effects of
modified gravity by updating these assignments to include a cell-centred value
of $G$, which differs from the Newtonian value in the region exterior to the
screening radius. This implementation uses a quasi-static approximation where,
given some initial radial profile, the star is evolved to its new equilibrium
structure one time-step later. Using this profile, we integrate equation
(\ref{eq:chiint}) to successively deeper cells until it is satisfied. The
radius of this cell is then designated the screening radius and we then update
the value of $G$ in each cell according to equation (\ref{eq:gprof}) so that
\begin{equation}
 G(r)=G\left[1+\alpha\left(1-\frac{M(\rs)}{M(r)}\right)\right]\quad r\ge \rs.
\end{equation}
We then let the model evolve one time-step further to find the modified
structure. This approximation is valid provided that the time-step between
successive models is smaller than the time-scale on which changes to $G(r)$ are
important and {\scriptsize MESA} provides a facility to ensure this is always
the case. Furthermore, \cite{Chang:2010xh} have modified {\scriptsize MESA} for
the same purpose of us using a scalar field ansatz and cell interpolation and
we have verified that our results are indistinguishable from theirs.

This modified version of {\scriptsize MESA} was presented and studied in
\cite{Davis:2011qf} and was used to make quantitative predictions of modified
Cepheid pulsation frequencies in \cite{Jain:2012tn}. More details can be found
in these references.

\section{Numerical Techniques}

Here we briefly describe the numerical methods used in order to obtain
solutions of the MLAWE. Whereas the form presented in equation
(\ref{eq:MLAWE}) is useful for extracting the new physical features,
numerically, it is more convenient to work with a first order system. One may
re-write the homogeneous MLAWE in the form
\begin{equation}\label{eq:1storderstart}
 \frac{\dd }{\dd r}\left[\frac{\gao P_0}{r^2}\frac{\dd}{\dd
r}\left(r^3\xi\right)\right]-4\frac{\dd P_0}{\dd
r}-4\pi\alpha G\rho_0^2r\xi+\omega^2r\rho_0\xi=0.
\end{equation}
In order to cast this into first order form we introduce the new variable
$\eta$, defined by
\begin{equation}
 \eta(r) = \frac{1}{r^2}\frac{\dd}{\dd
r}\left(r^3\xi\right),
\end{equation}
which, using equation (\ref{eq:lagpresspert}), is nothing but
$-\delta\rho/\rho_0$. Using this definition, we can cast
(\ref{eq:1storderstart}) into a convenient first-order from for both Lane-Emden
and {\scriptsize MESA} models.

\subsection{Lane-Emden Form}

When solving for Lane-Emden profile, we replace $r$ with $y$ using (\ref{eq:rc})
and the pressure and density with $\theta$ to obtain
\begin{align}
 \frac{\dd \eta}{\dd y} &= \frac{1}{\gao\theta}\left[4(n+1)\frac{\dd
\theta}{\dd
y}\xi+\alpha(n+1)y\theta^n\xi\right.\label{eq:leeq1}\\&\left.-(n+1)\gao\frac{\dd
\theta}{\dd y}\eta-\tilde{\omega}^2y\xi\right]\nonumber\\ \frac{\dd \xi}{\dd y}
&= \frac{\eta-3\xi}{y}\label{eq:leeq2},
\end{align}
where $\gao$ is treated as constant in Lane-Emden models, $\tilde{\omega}$
is defined in (\ref{eq:omtild}) and the term proportional to $\alpha$ is
present only when $r>\rs$. The value of $\gao$ may be chosen at will.
These are supplemented by the boundary conditions
\begin{align}
 \eta(0)&=3\xi(0)\label{eq:lecentrebc}\\
\eta(y_{\rm R}) &= \frac{\xi(y_{\rm
R})}{\gao}\left.\left(4+\frac{\tilde{\omega}^2y_{\rm R}}{(n+1)\dd\theta/\dd
y}\right)\right\vert_{y=y_{\rm R}},\label{eq:lesurfacebc}
\end{align}
which are the Lane-Emden equivalents of (\ref{eq:BCC}) and (\ref{eq:BCS}). We
solve this equation using the shooting technique. Using the fourth-order
Runge-Kutta method, we solve equations (\ref{eq:leeq1}) and (\ref{eq:leeq2}) in
conjunction with the modified Lane-Emden equation (\ref{eq:MGLE}) using a trial
value of $\tilde{\omega}^2$ and the centre boundary condition
(\ref{eq:lecentrebc}). We then test the surface condition
(\ref{eq:lesurfacebc}) against our solution, iterating over different
values of $\tilde{\omega}^2$ until the difference between our solution and the
boundary condition is less than some predefined tolerance (exactly how much
depends on the accuracy required for the eigenvalue). Using this method, one can
obtain $\tilde{\omega}^2$ and the corresponding eigenfunction to the desired
accuracy.

\subsection{MESA form}

{\scriptsize MESA} produces pressure, density, temperature etc. profiles in
physical units, and so it is convenient to use the dimensionless quantities
defined in (\ref{eq:dimlessP}), \ref{eq:dimlessrho}) and (\ref{eq:dimlessr}) so
that the MLAWE is
\begin{align}
 \frac{\dd \eta}{\dd
x}&=\left(\gao(x)\bar{P}_0(x)\right)^{-1}\left[-\eta\left(\bar{P}_0\frac{\dd
\gao}{\dd x}+\gao\frac{\dd \bar{P}_0}{\dd x} \right)\right.\\&+4\frac{\dd
\bar{P}_0}{\dd x}\left.\vphantom{-\eta\left(\bar{P}_0\frac{\dd
\gao}{\dd x}+\gao\frac{\dd \bar{P}_0}{\dd x} \right)}
+4\pi\alpha
x\bar{\rho}_0^2\xi-\Omega^2x\bar{\rho}_0\xi\right],\\\frac{\dd \xi}{\dd
x}&=\frac{\eta-3\xi}{x},
\end{align}
where, again, the term proportional to $\alpha$ is only present when $x>x_{\rm
s}\equiv \rs/R$ and $\Omega^2$ is the dimensionless eigenfrequency defined in
(\ref{eq:dimOmega}).

The eigenvalue problem is solved in a similar manner to the Lane-Emden models:
using profiles for $\bar{P}_0$ and $\bar{\rho}_0$ at both the cell boundaries
and centres as well as the screening radius for specified models from the
modified version of {\scriptsize MESA} described in Appendix A these equations
are integrated from the stellar centre using the fourth order Runge-Kutta scheme
for a test value of $\Omega^2$. $\Omega^2$ is then iterated until the boundary
condition at the stellar surface,
\begin{equation}
 \eta(1)=\frac{1}{\gao(1)}\left(\Omega^2+4\right)\xi(1),
\end{equation}
is satisfied up to some pre-set tolerance. In this manner, $\Omega^2$ and the
corresponding eigenfunction can be found for any given model and the period
found by inverting (\ref{eq:dimOmega}).
\bibliography{ref}
\end{document}